\documentclass[preprint,12pt]{elsarticle}

\usepackage{amssymb}
\usepackage{amsmath}
\usepackage{cite}
\usepackage{amsmath, amssymb, amsfonts}
\usepackage{graphicx}
\usepackage{textcomp}
\usepackage{xcolor}
\usepackage{bm}
\usepackage{enumerate}
\usepackage{threeparttable}
\usepackage{tabularx} 
\usepackage{float}
\usepackage{caption}
\usepackage{placeins}
\usepackage{latexsym}
\usepackage{subcaption}
\usepackage{wrapfig}
\usepackage{lscape}
\usepackage{pgfgantt}
\usepackage{times}
\usepackage{stfloats}
\usepackage{multirow}
\usepackage{xspace}
\usepackage{calc}
\usepackage{fancyhdr}
\usepackage{varioref}
\usepackage{tikz}
\usepackage{booktabs}
\usepackage{theorem}
\usepackage{longtable}
\usepackage{verbatim}
\usepackage{changebar}
\usepackage{algorithm}
\usepackage{algorithmicx}

\usepackage[noend]{algpseudocode} % Load algpseudocode package with noend option

\usepackage{url}
\usepackage{colortbl}
\usepackage{enumitem}
\usepackage{amsmath}
\usepackage{bm}
\usepackage{threeparttable}
\usepackage{makecell}
\usepackage{soul} % Required for \ul command
\usepackage{tabularx} % in the preamble
\usepackage{hhline}
\usepackage{boldline} 
\usepackage{tcolorbox}
\usepackage{mathtools}

\usepackage{appendix}  % For appendices
\usepackage{multirow}
\usepackage{pifont}% http://ctan.org/pkg/pifont

\newcommand{\algrule}[1][.2pt]{\par\vskip.5\baselineskip\hrule height #1\par\vskip.5\baselineskip}

{\theorembodyfont{\itshape} {\bfseries}{\rmfamily}}
{\theorembodyfont{\rmfamily}
{\bfseries}{\rmfamily}}
{\bfseries}{\rmfamily}
{\bfseries}{\rmfamily}
{\bfseries}{\rmfamily}

\newcommand{\specialcell}[2][c]{
\begin{tabular}[#1]{@{}c@{}}#2\end{tabular}}

\newcommand{\romannum}[1]{\uppercase\expandafter{\romannumeral #1\relax}}

\newcommand{\as}{\ensuremath {\leftarrow}{\xspace}}
\newcommand{\sk}{\ensuremath {sk}{\xspace}}

\newcommand{\pk}{\ensuremath {PK}{\xspace}}

\newcommand{\Ra}{\ensuremath \stackrel{\$}{\leftarrow}{\xspace}}

\newcommand{\sgn}{\ensuremath {\texttt{SGN}}{\xspace}}

\newcommand{\sgnsig}{\ensuremath {\texttt{SGN.Sig}}{\xspace}}
\newcommand{\sgnver}{\ensuremath {\texttt{SGN.Ver}}{\xspace}}
\newcommand{\sgnkg}{\ensuremath {\texttt{SGN.Kg}}{\xspace}}

\mathchardef\mhyphen="2D
\newcommand{\eucma}{\mathit{EU\mhyphen CMA}}

 \newtheorem{theorem}{Theorem}
{\theorembodyfont{\rmfamily}
{\theorembodyfont{\rmfamily}
\newtheorem{definition}{Definition}{\bfseries}{\rmfamily}}
{\theorembodyfont{\rmfamily}

\newcommand{\Attacker}{\mathcal{A}}

\newcommand{\hors}{\ensuremath {\texttt{HORS}{\xspace}}}
	\newcommand{\horskg}{\ensuremath {\texttt{HORS.Kg}}{\xspace}}
	\newcommand{\horssig}{\ensuremath {\texttt{HORS.Sig}}{\xspace}}
	\newcommand{\horsver}{\ensuremath {\texttt{HORS.Ver}}{\xspace}}

\newcommand{\bitmapremoverow}{\ensuremath {\texttt{REMOVE\_ROW}}{\xspace}}

\newcommand{\bitmapgetrowcol}{\ensuremath {\texttt{BITMAP\_GET\_ROW\_COLUMN}}{\xspace}}
\newcommand{\bitmapunsetindices}{\ensuremath {\texttt{BITMAP\_UNSET\_INDICES}}{\xspace}}

\newcommand{\bitmapextendmatrix}{\ensuremath {\texttt{BITMAP\_EXTEND\_MATRIX}}{\xspace}}
\newcommand{\bitmapcleanup}{\ensuremath {\texttt{BITMAP\_CLEANUP}}{\xspace}}

\newcommand{\mumhors}{\ensuremath {\texttt{MUM-HORS}{\xspace}}}
	\newcommand{\mumhorskg}{\ensuremath {\texttt{MUM-HORS.Kg}}{\xspace}}
	\newcommand{\mumhorssig}{\ensuremath {\texttt{MUM-HORS.Sig}}{\xspace}}
	\newcommand{\mumhorsver}{\ensuremath {\texttt{MUM-HORS.Ver}}{\xspace}}

\newcommand{\bitmap}{\ensuremath {\texttt{BM}{\xspace}}}

\newcommand{\indices}{{\textbf{indices}}}

\algrenewcommand\algorithmicend{\textbf{end}} % define \algorithmicend

\newcommand{\smallfalse}{\text{\texttt{False}}}
\newcommand{\smalltrue}{\text{\texttt{True}}}

\newcommand{\minusEquals}{\mathrel{\scalebox{0.75}[0.75]{$-$$=$}}}
\newcommand{\plusEquals}{\mathrel{\scalebox{0.75}[0.75]{$+$$=$}}}
\newcommand{\oplusEquals}{\mathrel{\scalebox{1}[1]{$\oplus$$=$}}}
\newcommand{\modPlusEquals}[1]{\stackrel{#1}{\plusEquals}}
\newcommand{\modMinusEquals}[1]{\stackrel{#1}{\minusEquals}}

\makeatletter
\newcommand{\algsep}[1]{%
  \par\noindent\hspace*{\dimexpr\algorithmicindent+10pt}%
  \rule[0.5ex]{0.35\columnwidth}{0.1mm}%
  \hspace{5pt}#1\hspace{5pt}%
  \rule[0.5ex]{0.35\columnwidth}{0.1mm}%
  \par\noindent%
}

\makeatother

\newcommand{\cmark}{\ding{51}}%
\newcommand{\xmark}{\ding{55}}%

\newcommand{\myalgrule}[1][1.3pt]{\par\vskip.5\baselineskip\hrule height #1\par\vskip.5\baselineskip}

\usepackage{tikz}

\newcommand{\lbbased}{\mathcal{LB}}
\newcommand{\hbbased}{\mathcal{HB}}

\journal{peer-reviewed journal}

\begin{document}

\begin{frontmatter}



%\title{Lightweight and Highly Key Utilized Multiple-Time PQ-Secure Digital Signature for Extended Authentication in NextG Networks} 

\title{Signer-Optimal Multiple-Time Post-Quantum Hash-Based Signature for Heterogeneous IoT Systems}

%% Article title


\author[AFFIL]{Kiarash Sedghighadikolaei \corref{cor1}} %% Author name
\ead{kiarashs@usf.edu}

\author[AFFIL]{Attila A. Yavuz} %% Author name
\ead{attilaayavuz@usf.edu}

\author[AFFIL]{Saif E. Nouma} %% Author name
\ead{saifeddinenouma@usf.edu}

\affiliation[AFFIL]{organization={Department of Computer Science and Engineering\\ University of South Florida},%Department and Organization 
            city={Tampa}, 
            state={FL},
            country={USA}}

\cortext[cor1]{Corresponding author}

\begin{abstract}

Heterogeneous Internet of Things (IoTs) harboring resource-limited devices like wearable sensors are essential for next-generation networks. Ensuring the authentication and integrity of security-sensitive telemetry in these applications is vital. Digital signatures provide scalable authentication with non-repudiation and public verifiability, making them essential tools for IoTs. However, emerging quantum computers necessitate post-quantum (PQ) secure solutions, yet existing NIST-PQC standards are costlier than their conventional counterparts and unsuitable for resource-limited IoTs. There is a significant need for lightweight PQ-secure digital signatures that respect the resource constraints of low-end IoTs.

We propose a new multiple-time hash-based signature called {\em Maximum Utilization Multiple HORS} (\mumhors) that offers PQ security, short signatures, fast signing, and high key utilization for an extended lifespan. \mumhors~addresses the inefficiency and key loss issues of HORS in offline/online settings by introducing compact key management data structures and optimized resistance to weak-message attacks. We tested \mumhors~on two embedded platforms (ARM Cortex A-72 and 8-bit AVR ATmega2560) and commodity hardware. Our experiments confirm up to 40$\times$ better utilization with the same signing capacity ($2^{20}$ messages, 128-bit security) compared to multiple-time HORS while achieving 2$\times$ and 156-2463$\times$ faster signing than conventional-secure and NIST PQ-secure schemes, respectively, on an ARM Cortex. These features make \mumhors~ideal multiple-time PQ-secure signature for heterogeneous IoTs.

\end{abstract}


% \begin{highlights}
% \end{highlights}

% \vspace{-2mm}
\end{frontmatter}

\section{Introduction} \label{sec:introduction}

Authentication and integrity are essential for safeguarding sensitive data in next-generation networked systems, enabling the growth of the Internet of Things (IoT) across sectors like healthcare \citep{ahad2020technologies}, military \citep{nguyen20216g}, and industry \citep{pradhan2020security}. In healthcare, wearable devices transmit critical medical data, such as heartbeats and blood sugar levels, where compromised information can adversely affect individual health and security, underscoring the necessity for robust data protection measures.

Digital signatures provide scalable authentication with non-repudiation and public verifiability and, therefore, offer a trustworthy authentication alternative for embedded medical settings \citep{Medauthex,Medauthex2,esemesem,INFHORS}. However, traditional signatures like RSA \citep{RSA} and ECDSA \citep{ECDSA} are resource-intensive (due to operations such as modular exponentiation and elliptic curve scalar multiplication), causing performance degradation, leading to issues like frequent battery replacements in embedded medical devices  \citep{ometov2016feasibility, ateniese2017low, Yavuz:2013:EET:2462096.2462108}. There are lightweight conventional alternatives often employing a pre-computation approach \citep{SCRA:Yavuz} to reduce computation but in favor of increased storage. Hence, optimizing storage and computational efficiency is essential for extending device utility. Besides performance hurdles, emerging quantum computers can break these conventional-secure signatures and their aforementioned lightweight variants via Shor’s algorithm \citep{Shor-algo}. Therefore, there is a critical need for PQ-secure digital signatures that can respect the resource limitations of low-end (embedded) IoT medical devices and applications.

NIST has standardized three classes of PQ-safe signatures \citep{nistCallProposals,darzi2023envisioning}: Lattice-based ($\lbbased$) CRYSTALS-Dilithium \citep{ducas2018crystals} and FALCON \citep{fouque2019fast}, and hash-based ($\hbbased$) SPHINCS+ \citep{SPHINCSPLUS}.  Dilithium is based on Fiat-Shamir with Aborts \citep{lyubashevsky2009fiat} and offers a compact public key size but with larger signatures. Falcon relies on a hash-and-sign scheme using NTRU lattices \citep{ducas2014efficient} but necessitates more complex operations and floating-point arithmetic. Both are more computationally intensive, with large keys and signatures compared to their conventional counterparts, making them unsuitable for highly resource-limited devices. For instance, Dilithium and Falcon-512 are 10-77$\times$ slower than ECDSA and SchnorrQ on ARM Cortex A-72 and require substantial storage (117KB for Falcon-512, 113KB for Dilithium on ARM Cortex-M4) \citep{kannwischer2019pqm4}. There are lightweight PQ-secure signatures that harness honest-but-curious verification servers with threshold control \citep{ANT:ACSAC:2021} and secure enclaves (e.g., Intel SGX \citep{Intel:SGX:2016}) for efficient verification and forward security \citep{HASESACMTOMM}. However, these special architectural and security assumptions may hinder their adaptation to some IoT applications as well as introduce potential security risks (e.g., secure enclave vulnerabilities \citep{secenclavevul}).

$\hbbased$ standards like XMSS \citep{XMSS} and SPHINCS+ \citep{SPHINCSPLUS} provide strong PQ security through hash functions with minimal assumptions. They build upon Few Time Signatures (FTSs) \citep{HORS,HORSIC,Winternitz:Sec:buchmann2011security} that permit efficient signing but only for a few messages. Stateful schemes like XMSS$^{MT}$ rely on Merkle Hash Trees (MHT) to enable Multiple-Time Signatures (MTS) from FTS, while stateless schemes such as SPHINCS+ use hypertree structures. Despite their merits, the signer overhead of these schemes is even higher than that of their lattice-based PQ-secure counterparts. Instead of general-purpose tree-based approaches, an alternative MTS is a conventional-secure public-key-based online/offline model. In this approach, public keys are pre-computed and stored at the verifier, while a hash-chain strategy is used at the signer with near-optimal efficiency~\citep{SEMECS,Yavuz:2012:TISSEC:FIBAF}. However, a straightforward transformation of FTSs such as HORS signatures \citep{HORS} into MTS via such an online/offline approach is shown to be inefficient \citep{SEMECS}. For instance, a HORS configuration with relatively short signatures leads to a key discard rate as high as 98\%, even without considering weak message security \citep{HORSWeakMessageAttack}. These high loss rates are detrimental to the life span and practicality of the target application. In Section \ref{sec:relatedwork}, we provide a more comprehensive revision of various alternative signatures and their pros and cons for low-end heterogeneous IoT applications. Overall, the state-of-the-art analysis suggests that {\em there is a vital need for lightweight $\hbbased$~signatures that permit multiple-time signature generation capability but with significantly better signer performance than existing PQ-secure alternatives}.

%%%%%%%%%%%%%%%%%%%%%%%%%%%%%%%%%%%%%%%%%%%%%%%%%%%%%%%%%% Contribution %%%%%%%%%%%%%%%%%%%%%%%%%%%%%%%%%%%%%%%%%%%%%%%%%%%%%%%%%%
\subsection{Our Contribution} \label{subsec:ourcontrib}
We created a new multiple-time hash-based signature referred to {\em Maximum Utilization Multiple HORS} (\mumhors), which is specifically signed for heterogeneous IoT applications with embedded/wearable medical devices as a representative use case. \mumhors~achieves PQ security with fast signing and compact signatures while permitting maximum public key utilization and key message security. Central to our scheme is a storage-efficient, fixed-size 2D bitmap for key management that facilitates efficient key derivation, public key management, signature failures, and weak-message security. We further outline the desirable features of our scheme as follows: \vspace{1mm}

$\bullet$~\ul{\em Fast Signing and Efficient Weak Message Resiliency:} The signing process of \mumhors~is similar to efficient HORS, offering fast signing, short signatures, and lower energy consumption. Additionally, we present an optimized mitigation method for weak message attacks \citep{HORSWeakMessageAttack} using XOR operations, which is significantly more efficient than techniques in HORSIC \citep{HORSIC}, HORSIC+ \citep{lee2021horsic+}, and PORS \citep{aumasson2018improving}. We evaluated the signing efficiency (cycles and energy consumption) of \mumhors~on various devices, including two embedded platforms (ARM Cortex A-72 and 8-bit AVR ATmega2560) and a commodity device. For instance, on the ARM Cortex A-72 (Table \ref{tab:embedded_comparison}), \mumhors~is faster by 24$\times$ than Falcon-512, by 200$\times$ than Dilithium, by 3900$\times$ than SPHINCS+, by 243-670$\times$ than XMSS variants, and even by 1.5-2$\times$ than conventional-secure schemes like ECDSA, Ed25519, and SchnorrQ. \vspace{1mm}

$\bullet$~\ul{\em Compact Key Management Data Structures:} Our key management data structure remains a small-constant size while offering a high capacity of signature generation. For instance, the size of our data structure is bounded as low as 1.4KB for a signing capacity of  $2^{20}$ signatures with 128-bit security, matching state-of-the-art multiple-time schemes like XMSS~\citep{XMSS:Buchmann:2011} and \citep{SEMECS,Yavuz:2012:TISSEC:FIBAF}. We provide a comprehensive theoretical and numerical analysis of the stability and accuracy of our data structures, ensuring they remain compact and practical even for highly constrained devices (e.g., 8-bit AVR ATmega2560). \vspace{1mm}

$\bullet$~\ul{\em Full-fledge Implementation:} \mumhors~full implementation is available at: 
\vspace{-5mm}
\begin{center}
\begin{tcolorbox}[colframe=black, colback=white, boxrule=0.1mm, sharp corners, boxsep=1mm, left=0mm, right=0mm, top=0mm, bottom=0mm, halign=center, text width=110mm]
\url{https://github.com/kiarashsedghigh/mumhors}
\end{tcolorbox}
\end{center}

{\em Potential Use Cases - Prioritizing Signer Optimality for Heterogeneous IoT Settings}: \mumhors~is a multiple-time signature designed for delay-tolerant and heterogeneous IoT applications that prioritizes signer efficiency and security (see Section \ref{sec:Models}). A typical application considered in state-of-the-art multiple-time signatures (e.g., \citep{SEMECS,Yavuz:2012:TISSEC:FIBAF,HASESACMTOMM}) is medical wearables (e.g., \citep{FitbitOfficial,AppleWatch}) and sensory devices in digital twins~\citep{aloqaily2022integrating}. In these use cases, the battery-powered IoT device regularly takes measurements, digitally signs them, and periodically uploads them to a resourceful cloud server. Ensuring battery longevity, minimal cryptographic resource usage and long-term PQ security of the low-end device is of utmost importance. To enable this, \mumhors, as in conventional-secure multiple-time alternatives  \citep{SEMECS,Yavuz:2012:TISSEC:FIBAF,HASESACMTOMM}, relies on an offline/online public-strategy that requires pre-computed public keys to be stored in a resource verifier (e.g., cloud server). As discussed further in Section \ref{sec:relatedwork}, this heterogeneous public-key model deviates from general-purpose signatures that rely on MHT or hypertrees (e.g., XMSS, SPHINCS) to avoid extreme burden on the signer in exchange for more storage on the cloud server. This is shown to be a highly favorable trade-off since a cloud server can maintain a larger public key to enable the optimized security and longevity of low-end devices in consideration, an aspect of heterogeneous computing in IoTs.

\begin{table}[ht!]
\centering
\caption{Comparison of Signature Schemes on ARM Cortex A-72 ($\kappa=128$-bit security)}

    \resizebox{\textwidth}{!}{
    \Huge
    \begin{tabular}{|c|c|c|c|c|c|c|c|c|c|c|c|c|c|c|c|}
    
    \hline
    \multirow{2}{*}{\textbf{Scheme}} & \multicolumn{2}{|c|}{\specialcell{\textbf{Signature Generation}}} & \multirow{2}{*}{\specialcell{\textbf{Private} \\ \textbf{key (KB)}}} & \multirow{2}{*}{\specialcell{\textbf{Signature} \\ \textbf{Size (KB)}}} & \multirow{2}{*}{\specialcell{\textbf{PQ} \\ \textbf{Promise}}} & \multirow{2}{*}{\specialcell{\textbf{Sampling} \\ \textbf{Operations}}} & \multirow{2}{*}{\specialcell{\textbf{Deterministic} \\ \textbf{Signing}}} 
    \\ \cline{2-3}

    & \textbf{8-bit AVR ATmega (cycles)} & \textbf{ARM Cortex A-72 ($\mu$s)} & & & & &  \\ \hline

    \multicolumn{8}{|c|}{\textit{Full-time Signatures ($M=2^{\kappa}$)}} \\ \hline
    ECDSA \citep{ECDSA} & 79 185 664 & 249.021 & 0.031 & 0.046 & \xmark & \xmark & \xmark  \\ \hline
    Ed25519 \citep{EdDSA} & 22 688 583 & 212.176 & 0.031 & 0.062 & \xmark & \xmark & \cmark \\ \hline
    SchnorrQ \citep{SchnorrQ} & 3 740 000 & 196.395 & 0.031 & 0.062 & \xmark & \xmark & \cmark \\

    \hline

    Falcon-512 \citep{fouque2019fast}  & N/A & 2 047.791 & 1.25 & 0.67 & \cmark & \cmark & \xmark \\ \hline
    Dilithium-II \citep{Dilithium} & N/A & 16 522.331 & 2.46 & 2.36 & \cmark & \cmark & \xmark  \\ \hline
    SPHINCS+ \citep{SPHINCSPLUS}     & N/A & 320 196.960 & 0.062 & 16.67 & \cmark & \xmark & \cmark \\ \hline

    \multicolumn{8}{|c|}{\textit{$M$-time Signatures ($M=2^{20}$)}} \\ \hline
    
    % Zaverucha et al. \citep{Zaverucha} &  &  & 0.017 & 0.046 & \xmark & \xmark & \cmark & \xmark \\ \hline
    SEMECS \citep{SEMECS} & 195 776 & 5.83 & 0.031 & 0.031 & \xmark & \xmark & \cmark\\
    \hline

    HORS \citep{HORS} & 342 976 & 46.8 & 0.031 & 0.78 & \cmark & \xmark & \cmark \\ \hline
    
    \multirow{2}{*}{$^{*}$HORSE \citep{HORSE}} & 342 976 & 46.69 & 790 MB & 0.078 & \multirow{2}{*}{\cmark} & \multirow{2}{*}{\xmark}  & \multirow{2}{*}{\cmark} \\
    & 4 979 776 & 682.52 & 480 & 0.078 & & & \\ \hline
    % LMS \citep{LMS} &  & & & \cmark & \xmark & \cmark & \xmark \\ 
    $^{**}$MSS \citep{rohde2008fast} & 5 792 000 & N/A  & 1.438 & 2.295 & \cmark & \xmark  & \cmark \\ \hline
    XMSS \citep{XMSS} & N/A & 20 943.975  & 1.34 & 2.44 & \cmark & \xmark & \cmark \\ \hline
    XMSS$^{MT}$ \citep{XMSSmt} & N/A & 55 099.507 & 5.86 & 4.85 & \cmark & \xmark & \cmark \\ \hline
    
    \hline

    \mumhors~ & 637 376 & 129.32 & 1.43 & 0.78 & \cmark & \xmark & \cmark \\ \hline
    % 342 976 + 294 400 #hash-cost(blake2b)=
    \end{tabular}
    }
    \begin{tablenotes}[flushleft] 
    			\scriptsize{ \item The parameter setting is as in Table \ref{tab:signature_comparison}. To the best of our knowledge, there is no reported benchmark on an 8-bit microcontroller for XMSS variants, Falcon-512, Dilithium-II, and SPHINCS+. $^{*}$It is impractical to implement HORSE variants on the AVR ATmega2560 MCU due to their large private key sizes. $^{**}$The maximum MSS signing capability is $2^{16}$ since the EEPROM memory endurance is less than $2^{20}$ write/erase cycles.  
       
       % \aay{This looks 8-bit from the table note writing, but it seems this is on Rasberry? There seems some confusion here. We must make the platform clear in this note.}
       }
    \end{tablenotes}
\vspace{-3mm}
\label{tab:embedded_comparison}
\end{table}

\section{Notations and Preliminaries} \label{sec:preliminaries}

\textbf{Notations:} $||$ and $|x|$ denote concatenation and the bit length of $x$, respectively. $x \Ra \mathcal{S}$ means $x$ is chosen uniformly at random from the set $\mathcal{S}$. $m \in \{0,1\}^*$ is a finite-length binary message. $\{q_i\}_{i=a}^{b}$ denotes $\{ q_a, q_{a+1}, \ldots, q_b\}$. $\log x$ is $\log_2x $. $[1,n]$ denotes all integers from $1$ to $n$. $f: \{0,1\}^* \rightarrow \{0,1\}^L$ and $H: \{0,1\}^* \rightarrow \{0,1\}^L$ denote one-way and cryptographic hash functions, respectively. $a * b$ and $a \stackrel{q}{*} b$ mean $a = a*b$ and $a = (a*b\mod q)$, where $*$ is an arbitrary operation.

% \textcolor{red}{make it generic SGN, not exclusively to hash, Def 4 will stand }
\vspace{-2mm}
%%% Hash-based digital signature %%%
\begin{definition} \label{def:sgn} A $\hbbased$ digital signature $SGN$ consists of three algorithms: 
 \begin{itemize}[leftmargin=8pt]
    \vspace{-1mm}
    \item[-] $\underline{ (\sk, \pk, I_{SGN}) \as \sgnkg( 1^{\kappa}) }$: Given the security parameter $\kappa$, it outputs the private key $\sk$, the public key $\pk$, and the system-wide parameters $I_{SGN}$.  
    \vspace{-3mm}
    \item[-] $\underline{ \sigma \as \sgnsig(\sk, m)  } $: Given the $\sk$ and message $m$, it returns a signature $\sigma$. 
    \vspace{-2mm}
    \item[-] $\underline{ b \as \sgnver(\pk, m, \sigma) }$: Given $\pk$, message $m$, and its corresponding signature $\sigma$,  it returns a bit $b$, with $b=1$ meaning valid, and $b=0$ otherwise.
	\end{itemize}
\end{definition}

\vspace{-3mm}
%%% HORS %%%
\begin{definition} \label{def:hors} Hash to Obtain Random Subset HORS \citep{HORS} is a $\hbbased$ digital signature consists of three algorithms:
\begin{itemize}[leftmargin=8pt]
    \vspace{-2mm}
    \item[-] $\underline{(\sk, \pk, I_\hors) \as \horskg(1^\kappa)}$: Given the security parameter $\kappa$, it selects $I_\hors \as (t, k, l)$, generates $t$ random $l$-bit strings $\{s_i\}_{i=1}^t$, and computes $v_i \as f(s_i), \forall i=1, \ldots, t$. Finally, it sets $\sk \as \{s_i\}_{i=1}^t$ and $\pk \as \{v_i\}_{i=1}^t$.
    \vspace{-3mm}
    \item[-] $\underline{\sigma \as \horssig(\sk, m)}$: Given \sk~and $m$, it computes $h \as H(m)$ and splits $h$ into $k$ $\log{t}$-sized substrings $\{h_j\}_{j=1}^k$ and interprets them as integers $\{i_j\}_{j=1}^k$. It outputs $\sigma \as \{s_{i_j}\}_{j=1}^k$.
    \vspace{-3mm}
    \item[-] $\underline{b \as \horsver(\pk, m, \sigma)}$: Given \pk, $m$, and $\sigma$, it computes $\{i_j\}_{j=1}^k$ as in $\horssig(.)$. If $v_{i_j}=f(\sigma_j), \forall j=1,\ldots,k$, it returns $b=1$, otherwise $b=0$.
	\end{itemize}
\end{definition}

\begin{definition} \label{def:hashrsrspr} Let $\mathcal{H}=\{H_{i,t,k,L}\}$ be a function family indexed by $i$, where $H_{i,t,k,L}$ maps an arbitrary length input to a $L$-bit subset of $k$ elements from the set $\{0, 1, ..., t-1\}$. $\mathcal{H}$ is $r$-subset (RSR) and second-preimage resistant (SPR), if, for every probabilistic polynomial-time (PPT) adversary $\Attacker$ running in time $\le T$: 

\vspace{-4mm}
\begin{equation*}
\begin{split}
\text{InSec}^{RSR}_{\mathcal{H}}(T) = \underset{\Attacker}{max}\{\text{Pr} & [(M_1, M_2, ..., M_{r+1}) \as {\Attacker}(i, t, k) \\ & \hspace{-27mm} \vspace{-19mm} \text{ s.t.}\; H_{i, t, k, L}(M_{r+1}) \subseteq \bigcup_{j=1}^{r}H_{i, t, k, L}(M_{j})]\} < \text{negl}(t,k)
\end{split}
\end{equation*}

\vspace{-5mm}

\begin{equation*}
\begin{split}
\text{InSec}^{SPR}_{\mathcal{H}}(T) = \underset{\Attacker}{max}\{\text{Pr} & [x \as \{0,1\}^*; x' \as {\Attacker}(x) \text{ s.t. } x \neq x' \\ & \hspace{-10mm} \text{and } H_{i,t,k,L}(x)=H_{i,t,k,L}(x')\} < \text{negl}(L)
\end{split}
\end{equation*}
\end{definition}

%%%%%%%%%%%%%%%%%%%%%%%%%%%%%%%%%%%%%%%%%%%%%% System model  %%%%%%%%%%%%%%%%%%%%%%%%%%%%%%%%%%%%%%%%%%%%%%
\section{Models and Use Cases} \label{sec:Models}
{\em \underline{System Model and Use-case}}: We assume a traditional public-key-based authentication model for heterogeneous IoT-cloud applications, wherein low-end IoT devices gather information in a store-and-forward manner for a delay-tolerant setting. Embedded healthcare devices such as pacemakers and implantable devices are prominent examples~\citep{yang2021efficient,salim2023lightweight}, in which the medical sensor takes continuous measurements (e.g., heart rate), digitally signs them, and then periodically uploads the telemetry and corresponding signatures into a cloud server for analysis. It is noted that other miscellaneous applications operate in similar settings, such as some smart cities \citep{lee2024slars} and drone services \citep{sc2023secure}. 

\begin{figure}[h!]
    \centering
    \includegraphics[width=\linewidth]{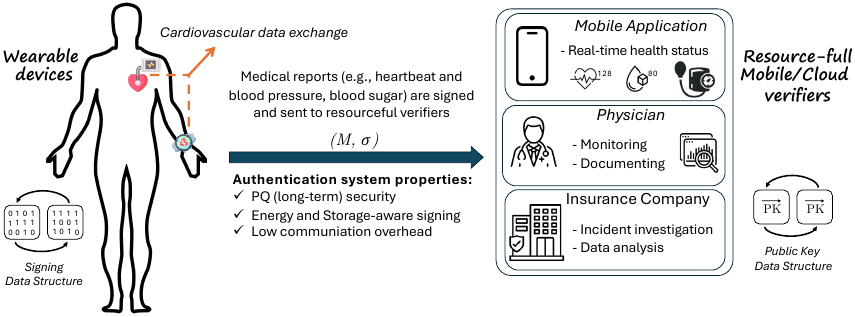}
    \caption{\mumhors~system model for a resource-limited wearable Medical IoT use-case.}
    \label{fig:mumhors_sys_model}
    \vspace{-1mm}
\end{figure}

In our target use case with wearable medical devices, the longevity, security, and efficiency of the low-end device are of primary concern. Specifically, our digital signature scheme focuses on energy-efficient computation (battery life and processing limits) and storage on resource-limited signers (e.g., 8-bit microcontrollers), while resourceful verifiers (e.g., cloud servers) manage public key storage and message authentication. Our system model is given in Figure \ref{fig:mumhors_sys_model}.

{\em \underline{Threat and Security Model}}: Our threat model assumes a quantum-computing capable adversary $\Attacker$ that can monitor all message-signature pairs and aims to intercept, modify, and forge them. The digital signature security model capturing our threat model follows the Existential Unforgeability under Chosen Message Attacks ($\eucma$).

\vspace{-2mm}
\begin{definition} \label{def:eucmasecurity} The $\eucma$ experiment for \sgn~is defined as follows:

\begin{itemize}[leftmargin=8pt]
    \vspace{-2mm}
    \item[-] $(\sk, \pk, I_{\sgn}) \,\as\, \sgnkg(1^\kappa)$ ,  
    $(m^*, \sigma^*) \,\as\, \Attacker^{\sgnsig_{sk}(.)}(\pk, I_{\sgn})$
    \vspace{-2mm}
    \item[-] $\Attacker$ wins the experiment after sending $q$ queries if $1 \as \sgnver(\pk, m^*, \sigma^*)$ and $m^*$ was not queried to the signing oracle $\sgnsig_{sk}(.)$.
    \vspace{-2mm}
\end{itemize}

\vspace{-1mm}
\begin{center}
    $Succ^{\eucma}_{\sgn}(\Attacker) = Pr[Expt^{\eucma}(\Attacker) = 1]$\\ \vspace{1.5mm}
    $InSec^{\eucma}_{\sgn}(T) = \underset{\Attacker}{max}\{Succ^{\eucma}_{\sgn}(\Attacker)\}$ $<$ $negl(T)$
\end{center}

\end{definition}
\section{Proposed Scheme} \label{sec:scheme}
{\em \underline{Main Idea}}: We propose \mumhors, a multiple-time hash-based signature scheme, which addresses severe key utilization limitations of HORS for multiple-time settings while ensuring maximum signer efficiency and weak message security. \mumhors~follows public-key offline-online model, in which public keys are pre-computed from a cryptographic hash chain approach (offline) and maintained at the verifier side, while the signer generates a signature from a master key efficiently (as in \citep{SEMECS}). However, as discussed in Section \ref{sec:introduction} and shown in~\citep{SEMECS}, this approach causes extremely large public keys and waste of individual components in public keys. This is due to the fact that, in HORS, to remain signer optimal by having efficient computation and communication, small values for $k$ enable efficient signature generation and short signatures. However, following the offline model results in the waste of significant $t-k$ keys. For instance, for the 128-bit security parameter setting ($t=1024$, $k=25$), only 2.44\% of public keys are effectively utilized, which poses a huge degradation to authentication usability and verifier storage. Moreover, the HORS signature is vulnerable to weak message attacks \citep{HORSWeakMessageAttack} as during signing, it does not guarantee the derivation of $k$ distinct index values ($i_j$), which causes the scheme to not meet the requirements of the r-subset-resiliency security (See Section \ref{sec:preliminaries}).

To address the outlined challenges, \mumhors~incorporates a storage-efficient two-dimensional bitmap with \( rt \) rows, each containing a \( t \)-bit list and metadata for correctness and optimization. During the signing, the first \( t \) set bits indicate available keys, with row and column indices aiding key derivation using the master key. Marked keys are used, while the remaining \( t-k \) keys stay unchanged, and depleted rows are replaced with new ones to maintain key availability. Moreover, \mumhors~includes an efficient weak key mitigation strategy (consisting of XOR operations) through algorithm design and parameter setup. 

{\em \underline{Outline of \mumhors~Algorithms}}: Algorithm \ref{alg:mumhors} outlines \mumhors, and Algorithm \ref{alg:bitmap} illustrates \bitmap~operations, as described below:

\begin{algorithm}[ht!]
    \footnotesize
    \caption{Maximum Utilization Multiple HORS (MUM-HORS)}\label{alg:mumhors}
    \hspace{5pt}
     
    %%%%%%%% Keygen %%%%%%%%
    \begin{algorithmic}[1]
        \vspace{-5pt}
        \Statex $ \underline{ (\sk, \pk, \bitmap, I_{\mumhors}) \as \mumhorskg(1^{\kappa}) } $:
        \vspace{-1pt}
        \State Set $I_{\mumhors} \as (I_{\hors}, r, rt)$, $msk \Ra \{0,1\}^\kappa$, and $\{pad_{i}\}_{i=1}^3 \Ra \{0,1\}^\kappa$

        \State Build bitmap $\bitmap = \{row_i\}_{i=1}^{rt}$ and set $row_i.num = i$, $row_i.activebits = t$, and set all $row_i.bits[.]$ to 1. Set $\bitmap.head = 1$, $\bitmap.tail = rt$, $\bitmap.window = t$, $\bitmap.nextrow = rt+1$, $\bitmap.activerows = rt$, and $\bitmap.activebits = rt\times t$

        \State Create public keys $\pk \!\!\as\!\! \{pk_{i}\}_{i=1}^r$ and set $pk_i.num = i$, $pk_i.activepks = t$, $\{pk_i.keys[j] \!\!\as\!\! f(PRF(msk || i || j))\}_{\substack{1 \leq i \leq r , 1 \leq j \leq t}}$. Set $\pk.head = 1$, $\pk.tail = rt$, $\pk.window = t$, $\pk.nextrow = rt+1$, $\pk.activerows = rt$, and $\pk.activepks = rt\times t$

        \State Send the private key $\sk \as msk$, $\{pad_{i}\}_{i=1}^3$, and $\bitmap$ to the signer, store the public keys $\pk$ and $\{pad_{i}\}_{i=1}^3$ on the verifier, and output the $I_{\mumhors}$
    \end{algorithmic}
    \myalgrule

    %%%%%%%%% Sign %%%%%%%%
    \begin{algorithmic}[1]
        \vspace{-5pt}
        \Statex $\underline{\sigma\as \mumhorssig(\sk, m)}$: Set $Ctr = 1$ and $hash \as Trunc(H(m), k\log t)$
        \vspace{3pt}

        % \algsep{\textit{Signer Online}}

        \State Split $hash$ into $k$ $\log t$-bit substrings $ \{ hash_j \}_{j = 1}^{k} $ and interpret each as an integer $i_j$
        \State \textbf{if} {$\exists p,q \in [1, k]$ s.t. $i_p = i_q$ and $p\neq q$} \textbf{then} $hash \oplusEquals pad_i$ and \textbf{goto} 1, $i \in [1,3]$, \textbf{else} \textbf{goto} 5

        % \State \textbf{if} no $pad_p$ was effective \textbf{then}:
        \State Split $hash' \as Trunc(H(hash || Ctr), k\log t)$ as step 1 into integer indices $i_j$ % into $k$ $\log t$-bit substrings $ \{ hash'_j \}_{j = 1}^{k} $ and interpret each as an integer $i_j$
        \State \textbf{if} {$\exists p,q \in [1, k]$ s.t. $i_p = i_q$ and $p\neq q$} \textbf{then} $Ctr \plusEquals 1$ and \textbf{goto} 3

        % \algsep{\textit{Signer Online}}
        \State Compute $\{(r_j, c_j)\} \as \bitmapgetrowcol(\bitmap, i_j)$, $\forall j \in [1, k]$    
        \State \Return  $\sigma \as (\{ PRF(sk || r_j || c_j) \}_{j = 1}^k, Ctr)$
        
        \algsep{\textit{Signer is Idle}}
        \State $\bitmapunsetindices(\bitmap, i_j)$, $\forall j \in [1, k]$
        \If {\bitmapextendmatrix(\bitmap) == $\smallfalse$} No more private keys to sign and \textbf{exit} \EndIf
    \end{algorithmic}
    \myalgrule

    %%%%%%%% Verify %%%%%%%%
    \begin{algorithmic}[1]
        \vspace{-5pt}
        \Statex $ \underline{ b \as \mumhorsver( \pk, m, \sigma)}$: Set $hash \as Trunc(H(m), k\log t)$%, $pre\_row=0$, $pre\_col=0$, and $b=1$
        \vspace{3pt}

        % \algsep{\textit{Verifier Online}}

        \State \textbf{if} $hash$ or any $pad_{1,2,3}$ yield distinct $i_j$ as steps 1-2 of $\mumhorssig(.)$, \textbf{then} \textbf{goto} 3 
        
        % \State Split $hash$ into $k$ $\log t$-bit substrings $ \{ hash_j \}_{j = 1}^{k} $ and interpret each as an integer $i_j$
        % \State \textbf{if} {$\exists p,q \in [1, k]$ s.t. $i_p = i_q$ and $p\neq q$} \textbf{then} $hash \oplusEquals pad_p$ and \textbf{goto} step 1, $p \in [1,3]$
        
        % \State \textbf{if} no $pad_p$ was effective \textbf{then} Set $hash' \as Trunc(H(hash || Ctr), k\log t)$ and:
        
        % \State \hspace{4mm} Split $hash'$ into $k$ $\log t$-bit substrings $ \{ hash'_j \}_{j = 1}^{k} $ and interpret each as an integer $i_j$

        \State Split $hash' \as Trunc(H(hash || Ctr), k\log t)$ as step 1 of $\mumhorssig(.)$ into integer indices $i_j$. \textbf{if} {$\exists p,q \in [1, k]$ s.t. $i_p = i_q$ and $p\neq q$} \textbf{then} Execute steps 3-4 of $\mumhorssig(.)$ and \textbf{return} $b=0$.

        \State Retrieve public key $pk_j$ for every index $i_j$ similar to $\bitmapgetrowcol(.)$ and if $(pk_j \neq f(\sigma_j))$ \textbf{return} $b=0$, $ \forall j  \in [1, k]$. Otherwise, \textbf{return} $b=1$
        
        \algsep{\textit{Verifier is Idle}}
        
        \State Invalidate all the $pk_j$ corresponding to the derived $i_j$ as in $\bitmapunsetindices(.)$
        \State Extend the view of the public keys similar to $\bitmapextendmatrix(.)$

        % \algsep{\textit{Verifier is Idle}}
        % \State Invalidate all the $pk_j$ corresponding to the derived $i_j$ as in $\bitmapunsetindices(.)$
        % \State Extend the view of the public keys similar to $\bitmapextendmatrix(.)$
    \end{algorithmic}
\end{algorithm}

\begin{algorithm}[ht!]
    % \footnotesize
    \caption{Bitmap Manipulation Algorithms}\label{alg:bitmap}
    \hspace{5pt} 
    \scriptsize
    %%%%%%%%%% BITMAP EXTEND %%%%%%%%%%
    \begin{algorithmic}[1]
        \vspace{-5pt}
	\Statex $b \as \underline{\bitmapextendmatrix (\bitmap)}$:
        % \vspace{-1pt}
        \If{$\bitmap.activebits < \bitmap.window$}
            \If{$\bitmap.nextrow \ge r$} \Return $\smallfalse$ \EndIf
            
            \If {$\bitmapcleanup(\bitmap) == 0$}
                \State Iterate over the rows and set $index$ to the index of the row with the minimum $.activebits$
                \State Update $\bitmap.activerows \minusEquals 1$, $\bitmap.activebits \minusEquals \bitmap[index].activebits$, and  $\bitmapremoverow(index)$
            \EndIf
            \State Set $fillcapacity = min(rt - \bitmap.activerows, r - \bitmap.nextrow)$. Add $fillcapacity$ rows where for each, update $\bitmap.tail \modPlusEquals{rt} 1$ and set $\bitmap[\bitmap.tail].num = \bitmap.nextrow$, $\bitmap[\bitmap.tail].activebits = t$, all $\bitmap[\bitmap.tail].bits[.]$ to 1, and $\bitmap.nextrow \plusEquals 1$. 
            
        \EndIf
            \State \Return $\smalltrue$ 
    \end{algorithmic}
    \myalgrule

    %%%%%%%%%% BITMAP CLEANUP %%%%%%%%%%
    \begin{algorithmic}[1]
	\Statex $n \as \underline{\bitmapcleanup(\bitmap)}$: Set $cleaned = 0$
        \State Iterate over the rows and \textbf{if} row $\bitmap[i].activebits == 0$ \textbf{then} $\bitmapremoverow(i)$ and $cleaned \plusEquals 1$

        \State \Return $cleaned$
    \end{algorithmic}
    \myalgrule
    
    %%%%%%%%%% BITMAP Get Row and Col %%%%%%%%%%
    \begin{algorithmic}[1]
        \Statex $(row, col) \as \underline{\bitmapgetrowcol(\bitmap, index)}$: %Set global $i=0$

        \State Iterate over the rows and \textbf{if}~{$index \ge \bitmap[i].activebits$} \textbf{then} $index \minusEquals \bitmap[i].activebits$ 
        \State Set $colidx$ as the index of $(index+1)^{th}$ set bit in $\bitmap[i].bits[.]$ and \textbf{return} $(\bitmap[i].num, colidx)$
    \end{algorithmic}
    \myalgrule
    
    \begin{algorithmic}[1]
        \Statex $\underline{\bitmapunsetindices(\bitmap, \indices)}$: 
        % \State Sort $\indices$ in descending order % and \textbf{if} {$\indices[i] \neq \indices[i-1]$} \textbf{then} $\indices[i] \minusEquals delta$ and $delta \plusEquals 1$ 
        \For{$index$ \textbf{in} $\indices$}

            \State Iterate over the rows and \textbf{if}~{$index \ge \bitmap[i].activebits$} \textbf{then} $index \minusEquals \bitmap[i].activebits$ 
            \State Unset the $(index+1)^{th}$ bit in $\bitmap[i].bits[.]$ to 0 and update $\bitmap.activebits \minusEquals 1$ {and} $\bitmap[i].activebits \minusEquals 1$
        \EndFor             
    \end{algorithmic}
    \myalgrule
    \begin{algorithmic}[1]
    \Statex $\underline{\bitmapremoverow(index)}$: 
        \If {$index == \bitmap.head$} $\bitmap.head = \bitmap.head \modPlusEquals{rt} 1$ \EndIf
        \State \textbf{else if} {$index == \bitmap.tail$} \textbf{then} $\bitmap.tail \modMinusEquals{rt} 1$
        \State \textbf{else}

        \State 
        \hspace{4mm} \textbf{if} {$\bitmap.head < \bitmap.tail$} \textbf{then}

            \State \hspace{8mm} \textbf{if}~{$index < \frac{\bitmap.tail - \bitmap.head}{2}$} \textbf{then}
                Shift rows from $\bitmap.head$ to $index$ and set $\bitmap.head \modPlusEquals{rt} 1$
            \State \hspace{8mm} \textbf{else}
                 \hspace{0.1mm} Shift from $\bitmap.tail$ to $index$ and set $\bitmap.tail \modMinusEquals{rt} 1$ 

        \State \hspace{4mm} \textbf{else}
            \State \hspace{8mm} \textbf{if} {$\bitmap.head < index$} \textbf{then}
                Shift rows from $\bitmap.head$ to $index$ and set $\bitmap.head \modPlusEquals{rt} 1$ \textbf{if}~{$index-\bitmap.head < rt - index$} \textbf{o.w.} 
                Shift rows from $\bitmap.tail$ to $index$ and set $\bitmap.tail \modMinusEquals{rt} 1$
            \State \hspace{8mm} \textbf{else}
                \hspace{0.1mm} Shift rows from $\bitmap.tail$ to $index$ and set $\bitmap.tail \modMinusEquals{rt} 1$ \textbf{if}~{$\bitmap.tail - index < index$} \textbf{o.w.} Shift rows from $\bitmap.head$ to $index$ and set $\bitmap.head \modPlusEquals{rt} 1$
                
    \end{algorithmic}
    % \myalgrule
    % \algfootnote

    % \begin{tablenotes}[flushleft] 
    % \tiny{ 
    % \item $\dag$ Note: Given the implementation as circular queue with $\bitmap.head$ and $\bitmap.tail$ pointing to the beginning and the end of the queue, when we iterate over, we consider the two cases where $\bitmap.head < \bitmap.tail$ and otherwise as follows:
    % \aay{Remove dashed line, space and technical notes title}
    % 1: \textbf{if}~{$\bitmap.head < \bitmap.tail$}\\
    % 2: \hspace{4mm} \textbf{for}~{$i = \bitmap.head$ to $\bitmap.tail$} // Do Something //\\
    % 3: \textbf{else}\\
    % 4:  \hspace{4mm} \textbf{for}~{$i = \bitmap.head$ to $rt$} \aay{not clear what this comment is, if this details are not of essence, but them in Appendix, this may confuse people} // Do Something // \\
    % 5:    \hspace{4mm} \textbf{for}~{$i = 0$ to $\bitmap.tail$} // Do Something // \\
    % }
    % \end{tablenotes}
        % \vspace{-3mm}

\end{algorithm}

%%%%%%%%%%%%%%%%%%%%%%%%%%%%%%%%%%%%%%%%%%%% Algorithm Description %%%%%%%%%%%%%%%%%%%%%%%%%%%%%%%%%%%%%%%%%%%%
The $\mumhorskg(.)$ initializes system parameters $I_{\mumhors}$, including the HORS and \bitmap~parameters $(r, rt)$, representing the total and maximum bitmap rows, respectively. It generates the master key $msk$ and three random pads (Step 1). An $rt$-row bitmap is created as a circular queue {(with middle node deletion support)} with $rt$ cells, initializing global parameters $head$ (first row's index), $tail$ (last row's index), $window$ (window size in bits), $nextrow$ (next row number), and $activerows$ (current number of rows). Row-specific parameters include $num$ (current row number), $activebits$ (number of active bits in the row), and $bits[.]$ (the $t$-bit list) (Step 2). Public keys are generated by concatenating $msk$ with each bit's row and column number (Step 3). The public key $\pk$ shares the bitmap parameters to ensure synchronization. The private key $sk$, pads $pad_{1,2,3}$, and \bitmap~are stored on the signer, while $\pk$ is stored on the verifier (Step 4).

The $\mumhorssig(.)$ first checks if the message hash produces distinct indices. If not, the hash is XORed with three pads; signing proceeds if any pad is effective (Steps 1-2). If none work, $Ctr$ is concatenated with the hash and iteratively hashed until $k$ distinct $\log t$-sized parts are obtained (Steps 3-4). The hash output is truncated to $k \log t$ bits for security (Steps 0 and 3). Once $k$ distinct indices are identified, row and column indices of the first $window$ set bits are retrieved from the \bitmap~using $\bitmapgetrowcol(.)$ (Step 5) to reconstruct the private keys as in key generation. The $\bitmapgetrowcol(.)$ (Algorithm \ref{alg:bitmap}) iterates over the bitmap rows to locate the $(index+1)^{th}$ set bit and returns its row and column indices. The iteration follows the circular queue iteration algorithm. Finally, the signature is generated (Step 6).

When the signer is idle after sending a signature, it unsets the derived $k$ indices in the bitmap using $\bitmapunsetindices(.)$ (Algorithm \ref{alg:bitmap}). Separating unsetting from index retrieval was a performance-driven decision. Private keys must correspond to the order of HORS indices derived from the message, requiring either a mapping data structure after sorting (required for unsetting) or separating these processes for efficiency. The indices are processed in descending order as computed in $\mumhorssig(.)$ (Steps 1-4), locating each index's position in the bitmap, setting the corresponding bit to 0, and updating the bitmap's global and row parameters (Steps 1-3). After updating, the signer checks if there are enough private keys for future messages by invoking $\bitmapextendmatrix(.)$ (Algorithm \ref{alg:bitmap}) (Step 8). This function extends the matrix if fewer than $window$ active bits are present. It first checks for and removes empty rows using $\bitmapcleanup(.)$ (Steps 2-3), which removes rows with zero active bits and returns the number of removed rows. If no rows are empty, the row with the fewest active bits is removed (Steps 4-6). Additional rows are then added to fill the \bitmap, initialized, and the bitmap parameters are updated (Step 6). If no extension was needed or it was successful, it returns $\smalltrue$ (Step 7).

$\mumhorsver(.)$ first verifies that the message hash and pads produce distinct indices (Step 1). If not, it checks the $Ctr$ and, in the case of an invalid $Ctr$, the verifier rejects the signature and computes a valid $Ctr$ to update the public key storage (Step 2). It then retrieves the public key corresponding to the ${i_j}^{th}$ public key within the first $window$ public keys. If any public key is rejected, the signature is deemed invalid; otherwise, the signature is accepted (Step 3). Finally, during idle periods, the verifier removes public keys from storage, similar to $\bitmapunsetindices(.)$, and if additional public keys are required, the verifier follows the $\bitmapextendmatrix(.)$~procedure (Steps 4-5). To address potential resynchronization (a common challenge in MTSs \citep{mcgrew2016state}) between the signer and verifier caused by transmission noise or adversarial corruption, we propose a mitigation algorithm (see \ref{sec:appendix-sca}) that helps with state recovery.

\section{Performance Analysis and Comparison} \label{sec:performance}
This section analyzes the performance of \mumhors~and its counterparts, focusing on signer storage overhead, key and signature sizes, and signing and verification times, with key generation occurring offline. We discuss the optimal selection of the row threshold ($rt$) in the bitmap and evaluate the computational complexity of each bitmap operation. Our analysis focuses on the implementation of \bitmap~as a circular queue with support for middle node deletion. An alternative linked-list version, detailed in \ref{sec:appendix}, enhances performance but requires additional memory for storing each node's successor address. We provide both analytical and detailed experimental analyses for the signer overhead on commodity hardware and two selected embedded devices.

\subsection{Parameter Selection and Experimental Setup}
The evaluation of \mumhors~and its counterparts on various devices has been conducted with a security parameter set to $\kappa = 128$-bit, as detailed in Tables \ref{tab:embedded_comparison}-\ref{tab:energy_cost}. 

\vspace{1mm}
\noindent {\em \underline{Hardware and Software Configurations}}: We used three types of devices for our evaluations.
% We performed the evaluation on a desktop with an Intel i9-11900K @ 3.5GHz processor and 64GB of RAM. For embedded devices, we selected an 8-bit ATmega2560 @ 16MHz microcontroller equipped with 256KB flash memory, 8KB SRAM, and 8 KB EEPROM. We also selected a Raspberry Pi 4 equipped with a Quad-core Cortex-A72 @ 1.8GHz with 8GB of RAM. 
First, we used a desktop with an Intel i9-11900K @ 3.5GHz processor and 64GB of RAM to evaluate the signature generation and verification performance. Second, to demonstrate the signature generation performance of \mumhors~for low-end (embedded) IoT settings, we used an 8-bit ATmega2560 @ 16MHz microcontroller with 256KB flash memory, 8KB SRAM, and 8KB EEPROM, as well as a Raspberry Pi 4 with a Quad-core Cortex-A72 @ 1.8GHz and 8GB of RAM. We only use a desktop to assess signature verification as our system model assumes a resourceful verifier. 
For the one-way and cryptographic hash functions (i.e., $f$ and $H$), we choose Blake2\footnote{https://github.com/BLAKE2/} due to its high efficiency on commodity hardware and low-end embedded devices. 
% Blake2 was implemented using the official GitHub repository \footnote{https://github.com/BLAKE2/}. 
% SHA2 (256/512 bits) was implemented using LibTomCrypt\footnote{https://github.com/libtom/libtomcrypt} and Blake2 using

\vspace{1mm}
\noindent {\em \underline{Parameter Selection}}: 
% The parameter selection for the counterparts follows their specifications. For \mumhors, we can derive the following based on the bitmap's two-dimensional structure given $M$ as the total number of messages to be signed:
The counterparts' parameters are set according to their 128-bit security specifications. For \mumhors, the parameter relationship can be derived from the bitmap's two-dimensional structure, with $M$ as the total number of messages to be signed and $r$ as the total required number of rows:

\begin{center}
$M$ = $(r-1)\cdot \frac{t}{k} + 1$
\end{center}

% The size of the bitmap data structure consists of $rt$ rows, where each row has $t$ bits, along with metadata of the row number and the number of active bits. Thus, the size of the bitmap is:
\bitmap~loads $rt$ rows at a time, each containing $t$-bit vector, along with metadata (row number and the number of active bits). Therefore, the size of the bitmap is:

\begin{center}
$|\bitmap|$ = $rt \cdot (t + \log{t} + \log{r})$ bits
\end{center}

The parameter $rt$ affects the bitmap size and private key usage (as well as the number of signable messages). Increasing $rt$ raises the chance of having rows with no active bits, as the remaining $t-k$ bits are distributed across more rows before invoking $\bitmapextendmatrix(.)$. This reduces key loss by permitting the replacement of empty rows through $\bitmapcleanup(.)$ rather than removing the row with fewer active bits. For example, with parameters $(t=1024, k=25, l=256, r=25601, rt=11, M=2^{20})$, the private key usage is 100\%, allowing all messages to be signed. However, setting $rt=8$ results in a loss of 10,536 bits and a reduction of 463 signable messages experimentally.

\vspace{1mm}
\noindent {\em \underline{Optimal Values for $rt$}}:
To derive the optimal bound for $rt$, we analyze the bitmap when the total number of bits is $t$, just before allocating a new row. The aim is to ensure that, with high probability, at least one row has fewer than $k$ active bits (ideally close to zero) so that key loss is minimized during row replacement when selecting $k$ bits among the rows. We formally define the problem as follows:

\noindent \underline{\textit{Question 1}}: {Given $rt$ rows and $t$ bits uniformly distributed among them, what is the minimum $rt$ such that, with high probability, at least one row has a maximum load of $k$ bits or less?}\label{def:rowproblem}

To answer this, we translate the problem into the Balls in Bins problem \citep{raab1998balls}. Specifically, given $m$ balls and $n$ bins, we aim to determine the maximum load in a bin with high probability after uniformly distributing the $m$ balls. For this purpose, we apply Theorem 1 from \citep{raab1998balls}:

{\theorem: Let $M$ be the random variable that counts the maximum number of balls in any bin. We throw $m$ balls independently and uniformly at random into $n$ bins. Then with high probability, $M > {load}_{max}$ and we have:}

\vspace{-5mm}
\begin{equation}
load_{max}=\frac{m}{n} + \sqrt{2 \frac{m \log n}{n} \left(1 - \frac{1}{\alpha}\frac{\log \log n}{2 \log n} \right)}, \text{if } m \gg n \cdot (\log n)^3, 0<\alpha<1
\end{equation}

In the above context, $m$ balls correspond to $t$ bits, $n$ bins correspond to $rt$ \bitmap~rows, and $\alpha$ is the smoothing parameter. {Increasing \(\alpha\) provides a more conservative estimate of the maximum load, ensuring that, with high probability, the maximum load is almost the estimated bound. In essence, a higher \(\alpha\) shifts the uniform ball distribution \(\frac{m}{n}\) towards \(load_{max}\).} While this approach determines an upper bound on the maximum load, we are interested in having $\le k$ bits in at least one row with high probability (minimizing the load). Thus, we define:

\noindent \underline{\textit{Dual of Question 1}}: Given a bitmap with $rt \cdot t$ bits, where $t$ bits are present, and $rt \cdot t - t$ bits are marked as used, minimizing the number of unmarked bits in at least one row is equivalent to maximizing the number of marked bits in at least one row. What should $rt$ be to ensure that, with $rt \cdot t - t$ marked bits added, the maximum load in at least one row is $t-k$ or ideally $t$?

To solve this, we set $ m = rt \cdot t - t $, $n=rt$ and $load_{max}=t-k$ and we get:
\begin{equation}
\begin{aligned}
t-k &= \frac{rt \cdot t - t}{rt} + \sqrt{2 \frac{(rt \cdot t - t) \log rt}{rt} \left(1 - \frac{1}{\alpha} \frac{\log \log rt}{2 \log rt} \right)}, \\
&\quad \text{if } rt \cdot t - t  \gg rt \cdot (\log rt)^3, \quad 0 < \alpha < 1
\end{aligned}
\end{equation}

We analyze the parameters by implementing a numerical solver (in MATLAB, available in our code repository\footnote{\url{https://github.com/kiarashsedghigh/mumhors/Optimizations/row_threshold.m}}). With parameters $t=1024$ and $k=25$ and setting $\alpha=0.999$, our solver outputs a row threshold of 10.903, assuming a maximum load of $t-k$. To increase the likelihood of full depletion, assuming $load_{max}=t$, the solver outputs a safer margin of 13.94 rows.

\subsection{Efficiency Evaluation and Comparison}
Given the performance evaluation on two embedded platforms and a commodity device in Tables {\ref{tab:embedded_comparison}-\ref{tab:energy_cost}}, takeaways are as follows:

\begin{table}[H]
    \centering
    \caption{Comparison of Signature Schemes on Commodity Hardware ($\kappa=128$-bit security)}
    \resizebox{\textwidth}{!}{
    \Huge
    \begin{tabular}{|c|c|c|c|c|c|m{3.5cm}<{\centering}|c|c|c|}
        \hline
        \multirow{2}{*}{\textbf{Scheme}} & \multirow{2}{*}{\textbf{\specialcell{Sig Gen \\ (\bm{$\mu$}s)}}} & \multirow{2}{*}{\textbf{\specialcell{Sig Size \\ (KB)}}} & \multirow{2}{*}{\textbf{\specialcell{sk Size \\ (KB)}}} & \multirow{2}{*}{\textbf{\specialcell{Sig Ver \\ (\bm{$\mu$}s)}}} & \multirow{2}{*}{\textbf{\specialcell{PK Size \\ (KB)}}} & \multirow{2}{*}{\textbf{\specialcell{PQ \\ Promise}}} & \multirow{2}{*}{\textbf{\specialcell{Memory Exp \\ Code Size}}} & \multirow{2}{*}{\textbf{\specialcell{Sampling \\ Operations}}} & \multirow{2}{*}{\textbf{\specialcell{Deterministic \\ Signing}}} \\
         & & & & & & & & & \\ \hline

        \multicolumn{10}{|c|}{\textit{Full-time Signatures ($M=2^{\kappa}$)}} \\ \hline
        ECDSA \citep{ECDSA} & 15.81 & 0.046 & 0.031 & 46.24 & 0.062 & \xmark & L & \xmark & \xmark \\ 
        Ed25519 \citep{EdDSA} & 12.14 & 0.062 & 0.031 & 33.17 & 0.031 & \xmark & L & \xmark & \cmark \\ 
        SchnorrQ \citep{SchnorrQ} & 8.57 & 0.062 & 0.031 & 15.42 & 0.031 & \xmark & L & \xmark & \cmark \\ 

        \hline \hline
        Falcon-512 \citep{fouque2019fast} & 170.45 & 0.67 & 1.25 & 28.21 & 0.87 & \cmark & M & \cmark & \xmark \\ 
        Dilithium-II \citep{Dilithium} & 243.07 & 2.36 & 2.46 & 59.05 & 1.28 & \cmark & M & \cmark & \xmark \\ 
        SPHINCS+ \citep{SPHINCSPLUS} & 5217.01 & 16.67 & 0.062 & 415.41 & 0.031 & \cmark & H & \xmark & \cmark \\ 
        
        \hline 
        \multicolumn{10}{|c|}{\textit{$M$-time Signatures ($M=2^{20}$)}} \\ \hline
        Zaverucha et al. \citep{Zaverucha} & 14.97 & 0.046 & 0.031 & 28.2 & 4.06GB & \xmark & L & \xmark & \xmark \\ 
        SEMECS \citep{SEMECS} & 0.68 & 0.031 & 0.031 & 11.9 & 96MB & \xmark & L & \xmark & \cmark \\ 
        \hline \hline
        HORS \citep{HORS} & 6.12 & 0.78 & 0.031 & 6.35 & 32GB & \cmark & L & \xmark & \cmark \\ 
        \multirow{2}{*} {HORSE \citep{HORSE}} & 6.19 & \multirow{2}{*}{0.78} & 790 MB & 6.41 & \multirow{2}{*}{32} & \multirow{2}{*}\cmark & \multirow{2}{*}H & \multirow{2}{*}\xmark & \multirow{2}{*} \cmark \\ & 86.25 &  & 480 & 6.46 & & & & &  \\

        XMSS \citep{XMSS} & 2778.1 & 2.44 & 1.34 & 751.17 & 0.062 & \cmark & H & \xmark & \cmark \\

        XMSS$^{MT}$ \citep{XMSSmt} & 6785.85 & 4.85 & 5.86 & 1297.94 & 0.062 & \cmark & H & \xmark & \cmark \\ \hline\hline
        \mumhors & 17.56 & 0.78 & 1.43 & 19.17 & 800MB & \cmark & L & \xmark & \cmark \\ \hline
    \end{tabular}
    }
\label{tab:signature_comparison}
    \begin{threeparttable}
    \begin{tablenotes}[flushleft] 
    \scriptsize{
    \item The message size is 32 bytes. In the memory expansion and code size column, H denotes high, M denotes medium, and L denotes low. For SPHINCS+, parameters are $(n, h, d, t, k, w) = (16, 66, 22, 64, 33, 16)$. For XMSS, XMSS-SHA2\_20\_256 was selected, and for XMSS$^{MT}$, XMSSMT-SHA2\_{20}/2\_{256} was used. HORS and HORSE parameters are $(t, k, l) = (1024, 25, 256)$, with $d$ for HORSE set to 25290, where $d = M \cdot (1 - e^{-k/t})$. HORSE can reduce memory by using more hash computations (Jakobsson \citep{jakobsson2002fractal} method), which, while decreasing private key size, increases signing time. HORS has been converted to an M-time signature using the same method as \mumhors. For \mumhors, parameters are $(t, k, l, r, rt) = (1024, 25, 256, 25601, 11)$. The public key size for \mumhors~is 25600 full rows plus one row with 25 keys, totaling 800.00076 MB. The $f(.)$ that is used for HORS, HORSE, and \mumhors~is Blake2-256.}

    \end{tablenotes}
    \end{threeparttable}
\end{table}

\begin{table}[ht!]
\centering
\caption{Energy Usage of Signature Generation and Transmission on an AVR ATMega2560 MCU}

    \resizebox{\textwidth}{!}{
    \Huge
    \begin{tabular}{|c|c|c|c|c|c|c|c|c|c|c|c|c|c|c|c|}
    
    \hline
    \multirow{2}{*}{\textbf{Scheme} }
    & \multicolumn{2}{c|}{\specialcell{\textbf{Signature Generation} } }
    & \multicolumn{2}{c|}{ \specialcell{\textbf{Signature Transmission} } } 
    & \multirow{2}{*}{ \specialcell{\textbf{Total Energy} \\ \textbf{Cost (mJ)}}}
    & \multirow{2}{*}{ \specialcell{\textbf{Max Signing} \\ \textbf{Operations}}}
    & \multirow{2}{*}{\specialcell{\textbf{Private} \\ \textbf{key (KB)}}} 
    & \multirow{2}{*}{\specialcell{\textbf{PQ} \\ \textbf{Promise}}}  \\ \cline{2-5}
    
    & \textbf{Time (cycles)}
    & \textbf{Energy Cost (mJ)}
    & \textbf{Sig Size (KB)}
    & \textbf{Energy Cost (mJ)}
    &
    &
    &
    & \\ \hline

    \multicolumn{9}{|c|}{\textit{Full-time Signatures ($M=2^{\kappa}$)}} \\ \hline
    ECDSA \citep{ECDSA} & 79 185 664 & 332.285 & 0.046 & 0.065 & 332.35 & 20 316 & 0.031 & \xmark \\ \hline
    Ed25519 \citep{EdDSA} & 22 688 583 & 92.343 & 0.062 & 0.086 & 92.429 & 73 063 & 0.031 & \xmark \\ \hline
    SchnorrQ \citep{SchnorrQ} & 3 740 000 & 15.222 & 0.062 & 0.086 & 15.308 & 440 946 & 0.031 & \xmark \\
    \hline

    \multicolumn{9}{|c|}{\textit{$M$-time Signatures ($M=2^{20}$)}} \\ \hline
    
    SEMECS \citep{SEMECS} & 195 776 & 0.797 & 0.031 & 0.043 & 0.84 & 8 035 714 & 0.031 & \cmark \\ \hline
    HORS \citep{HORS} & 342 976 & 1.396 & 0.78 & 1.075 & 2.471 & 2 731 688 & 0.031 & \cmark \\ \hline
    $^{*}$MSS \citep{rohde2008fast} & 5 792 000 & 23.573  & 2.295 & 29.964 & 53.537 & 126 081 & 1.438 & \cmark \\ \hline
    \hline
    \mumhors~ & 637 376 & 2.594 & 0.78 & 1.075 & 3.669 & 1 839 738 & 1.43 & \cmark \\ \hline
    % 342 976 + 294 400 #hash-cost(blake2b)=
    \end{tabular}
      }
    \begin{tablenotes}[flushleft] 
    			\scriptsize{ \item The parameter setting is as in Table \ref{tab:signature_comparison}.

       \item $^{*}$The maximum MSS signing capability is $2^{16}$ since the EEPROM memory endurance is less than $2^{20}$ write/erase cycles.
       }
    \end{tablenotes}
\label{tab:energy_cost}
\end{table}

\vspace{2pt}
\noindent {\em \underline{Signer and Verifier Memory Usage}}: 
 The \mumhors~signer storage includes a 256-bit master key and a 1.4KB bitmap. The total storage overhead of 1.43KB is comparable to Falcon-512 and XMSS and is 1.7$\times$ smaller than Dilithium-II and 4$\times$ than XMSS$^{MT}$. Compared to HORSE, which uses hash chains, \mumhors~offers 335-500,000$\times$ memory savings. \mumhors~has low memory expansion and code size compared to PQ standards. On an ARM Cortex-M4, Falcon-512 needs 117KB, Dilithium 113KB, and SPHINCS+ 9KB for signing and verifying \citep{kannwischer2019pqm4}. 

\mumhors~verifier storage is 5$\times$ smaller than Zaverucha et al. \citep{Zaverucha} and 40$\times$ smaller than HORS and deletes the public keys from the memory as they are used.

\vspace{2pt}
\noindent {\em \underline{Communication Bandwidth Overhead}}:
 Smaller $k$ values in \mumhors~reduce the signature size with fewer private keys. The signature size is comparable to Falcon-512, 3$\times$ smaller than Dilithium-II, and 21$\times$ smaller than SPHINCS+. \mumhors~has similar signature size as HORS and HORSE, is 6.21$\times$ smaller than XMSS$^{MT}$, but is larger than non-PQ secure SEMECS, Zaverucha et al. \citep{Zaverucha}, and full-time ECDSA, Ed25519, and SchnorrQ counterparts.

\vspace{2pt}
\noindent {\em \underline{Signing on the Commodity Device and Verification}}:
\mumhors~achieves fast signing and verification. Signing is 17$\times$, 24$\times$, and 511$\times$ faster than Falcon-512, Dilithium-II, and SPHINCS+, respectively, and 1.5$\times$ and 1.2$\times$ faster than ECDSA and EdDSA, respectively. While performance is comparable to ECDSA, Ed25519, and \citep{Zaverucha}, \mumhors~is 665$\times$ faster than XMSS$^{MT}$. Verification speed matches signing speed, making \mumhors~ 118$\times$ and 38$\times$ faster than PQ-secure XMSS$^{MT}$ and SPHINCS+, respectively, and 4.5$\times$ and 3$\times$ faster than ECDSA and EdDSA. Moreover, $\bitmapcleanup(.)$ takes 0.027$\mu s$, $\bitmapunsetindices(.)$ 0.18$\mu s$ per index, and $\bitmapgetrowcol(.)$ 0.23$\mu s$ on commodity hardware, with the latter being the only one active during signature generation. It should be noted that $\bitmapextendmatrix(.)$ does not always remove rows as it is conditioned on the presence of $window$ bits in the \bitmap.

\vspace{2pt}
\noindent \underline{\em Signature Generation on Embedded Devices}:
The efficiency of \mumhors~on embedded devices (ARM Cortex A-72) has been presented in Table \ref{tab:embedded_comparison}. On the ARM Cortex A-72, \mumhors~signing is 1.75$\times$ slower than its baseline HORS due to bitmap operations required for achieving a 41-fold public key size reduction. NIST finalists, like Dilithium, show significant delays in signing (e.g., 16 seconds), making them unsuitable for real-time applications. Moreover, operations $\bitmapcleanup(.)$ takes 0.6$\mu s$, $\bitmapunsetindices(.)$ takes 0.89$\mu s$ per index, and $\bitmapgetrowcol(.)$ takes 1.53$\mu s$.

\vspace{2pt}
\noindent \underline{\em Energy Impact and Cost of Signing on Embedded devices}: 
In Table \ref{tab:energy_cost}, we present the analysis of the impact of signature generation on energy consumption and battery life in 8-bit AVR microcontrollers using the energy estimation model from \citep{piotrowski2006public} based on the MICAz sensor node. The MICAz node features an ATmega128L microcontroller (16 MHz, 4KB EEPROM, 128KB flash memory) and a ZigBee 2.4 GHz radio chip (CC2420). It is powered by AA batteries with a maximum energy capacity of 6750J. According to \citep{piotrowski2006public}, the ATmega128L MCU consumes $4.07$nJ per cycle, while the CC2420 transceiver chip draws $0.168$mJ per bit transmitted. Using these results, we estimated the energy consumption for signing and transmission for \mumhors~and its counterparts. 

The signing capability depends on both signature generation and transmission efficiency. Notably, \mumhors~can support a larger number of signatures ($2^{20}$) than the practical limit of signing operations achievable on a MICAz node ($\approx 2 \cdot 10^6$). 
Our $\hbbased$ counterpart, the Merkle Signature Scheme (MSS) \citep{rohde2008fast}, supports only $\frac{1}{14}$ of the signatures achievable with \mumhors. MSS uses Winternitz One-Time Signature (WOTS) and constructs a Merkle tree, as in XMSS \citep{XMSS}, leading to higher signing costs due to the WOTS signature process and path computation in the tree.
Our second $\hbbased$ counterpart, HORS, offers $1.48\times$ more signatures than \mumhors, but at the cost of a significantly larger public key, approximately $40\times$ larger.
The elliptic-curve-based SEMECS can generate more signatures than the MICAz node's practical limit. However, it does not provide post-quantum security. 
\mumhors~offers the optimal balance between maximum signing capability, post-quantum security, and compact key sizes.

\section{Security Analysis} \label{sec:security}

\begin{theorem} 
\mumhors~is $\eucma$ secure if $H()$ is $r$-subset-resilient and second-preimage resistant:

\vspace{-3mm}
\begin{equation*}
\text{InSec}^{{\eucma}}_{\mumhors}(T) = \text{InSec}^{\text{RSR}}_{H}(T) + \text{InSec}^{\text{SPR}}_{H}(T) < \text{negl}(t, k, L)
\end{equation*}
\end{theorem}

\noindent \textit{Proof:} Given a set of adaptively chosen and queried $q$ valid message-signature pairs $\{(m_i, \sigma_i)\}^q_{i=1}$, there are the below cases where the adversary $\Attacker$ can forge a signature: 

\begin{itemize}[leftmargin=8pt]
    \item[-] \underline{\em $\Attacker$~breaks $r$-subset-resilient of $H$}: The $\Attacker$ identifies a message $m^*$ such that its $k$ distinct elements (as in $\mumhorssig(.)$) are among the observed $q \cdot k$ elements from the last $q$ messages, with a success probability of $(\frac{q \cdot k}{t})^k$. We note that our efficient mitigation method against weak message attacks ensures the derivation of $k$ distinct indices from a message, either by using its initial hash XORed with random pads secured by long-term certificates on the verifier, or through an iterative procedure using an incremental counter $Ctr$.

    The success probability $(\frac{q \cdot k}{t})^k$ decreases to $(\frac{k}{t})^k$ due to the bitmap design. Once the signer selects and uses $k$ bits from the first window, they are marked as used, preventing reuse in future rounds. Uniqueness is maintained by the row parameter $num$, ensuring each bit's combination $(msk || row || col)$ is distinct. Even when $\bitmapextendmatrix(.)$ is invoked, new rows have unique row numbers, enforced by the global $nextrow$ parameter. Moreover, the verifier maintains identical global and row parameters to manage public keys, ensuring synchronization with the signer. This guarantees that once a private key is used, the corresponding public key is invalidated during verification, preventing reuse. 
    
    In summary, since the bitmap replaces $k$ out of $t$ private keys, and the remaining $t-k$ unused keys are independent and hidden from the attacker, the success probability per round is reduced to that of HORS as an OTS, which is negligible for suitable $k$ and $t$. This probability of forging a signature on HORS, in addition to $(\frac{q \cdot k}{t})^k$, entails the likelihood of inverting the one-way function $f(.)$ to derive private keys from the verifier's stored public keys. Grover's algorithm \citep{Grover:1996} can reverse a black-box function with input size \(N\) in \(O(\sqrt{N})\) steps and \(O(\log_2 N)\) qubits. For $L$-bit output $f(.)$, the probability of reverting $k$ public keys is $2^{-k \cdot \frac{L}{2}}$.

    \item[-] \underline{\em $\Attacker$~breaks the second-preimage resistance of $H$}: We evaluate the security of our hash function $H(.)$ using Grover's model. The attacker could find $m^*$ such that $H(m_i) = H(m^*)$ and produce a valid $(m^*, \sigma_i)$. Given identical hashes, the $k$ indices for $m^*$ will match those for $m_i$ by any method (initial hash, pads, or $Ctr$). For an $L=256$-bit hash function like Blake2-256, the collision probability is $\frac{1}{2^{\frac{L}{2}}}$, providing 128-bit security, which is negligible. Moreover, the parameters $k$ and $t$ impact the security of $H(.)$, requiring $k\log t = L$. If $k\log t < L$, the attacker's success probability increases to $\frac{1}{2^{\frac{k\log t}{2}}}$. To maintain security, we ensure $k\log t = L$ by truncating hash outputs to $k\log t$ bits using the $Trunc(.)$ function during the signing.

    Overall, we conclude with an upper bound:
    \vspace{-3mm}
    \begin{equation*}
    \text{InSec}^{{\eucma}}_{\mumhors}(T) <  max(\frac{1}{2^{\frac{L}{2}}}, \frac{1}{2^{\frac{k\cdot \log {t}}{2}}}) + 2^{-k \cdot \frac{L}{2}} + (\frac{k}{t})^k 
    \end{equation*}
    
    We set the length of the master key ($|msk|$) and private key (HORS $l$ parameter) to $\kappa$ bits to ensure the minimum required security.

\end{itemize}
\section{Related Work} \label{sec:relatedwork}

\noindent {\em \textbf{Symmetric-key based Approaches}}: Two primary schemes, namely HMAC \citep{ModernCryptoBellareRogaway} and symmetric ciphers \citep{StinsonCrypto} are commonly used in IoT systems for their computational efficiency \citep{Yavuz:EncryptionMode:Patent}. However, these approaches lack scalability in large, distributed systems and do not provide public verifiability, which is essential for health audits \citep{halperin2008security} and non-repudiation in legal contexts \citep{camara2015security}. Digital signatures \citep{ECDSA,Ed25519,RSA}, though less efficient computationally, support scalable, publicly auditable authentication.

%%%%%%%%%%%%%%%%%%%%%%%%%%%%%%%%%%%%%%%%%%%%%%%%%%%%%%%%%% Conventional Signatures %%%%%%%%%%%%%%%%%%%%%%%%%%%%%%%%%%%%%%%%%%%%%%%%%%%%%%%%%%
\noindent {\em \textbf{Conventional Digital Signatures}}: Conventional digital signatures like ECDSA \citep{ECDSA}, Ed25519 \citep{Ed25519}, RSA \citep{RSA}, and BLS \citep{BLS:2001} are considered for IoT  \citep{pawar2023optimization}, wireless spectrum management~\citep{Yavuz:PrivacyCognitiveRadio:WSCNIS:Cuckoo:2015}, 6G \citep{kazmi2023security},  and various other network domains. However, security protocols relying on these primitives have been shown to deplete battery life in resource-constrained devices and may require frequent undesirable maintenance~\citep{ometov2016feasibility,ateniese2017low}. For example, while RSA is efficient in verification, it requires expensive signing operations and has large key sizes, and although ECDSA improves efficiency with smaller keys, it still incurs high computational costs. Moreover, relying on traditional intractability assumptions leaves these schemes vulnerable to quantum attacks like Shor's algorithm \citep{Shor-algo}, with ECC-based ones being more susceptible than RSA \citep{bernsteinRSAPostPQC,QuantumCircuitDLP,FactorRSA,proos2003shor}. \vspace{1mm}

%%%%%%%%%%%%%%%%%%%%%%%%%%%%% Lightweight Conventional Signatures with Special Trade-offs %%%%%%%%%%%%%%%%%%%%%%%%%%%%%
\noindent {\em \textbf{Lightweight Conventional Signatures with Special Trade-offs}}: Efforts to develop lightweight digital signatures that improve conventional methods like ECDSA and SchnorrQ often rely on advanced pre-computation and storage techniques, as done by SCRA \citep{SCRA:Yavuz} and Nouma et al. in~\citep{10.1145/3576842.3582376}. These schemes rely on the online/offline signature paradigm~\citep{shamir2001improved} involving pre-computed tokens. For instance, Rapid Authentication (RA) \citep{Yavuz:RA:TIFS:2014:Real.Time.BroadAuth} leverages pre-computed token aggregation for fast online signature generation. SEMECS \citep{SEMECS} optimizes EC-based signature schemes by reducing signature and private key sizes of \citep{Yavuz:2013:EET:2462096.2462108}, addressing the computational burden of deriving ephemeral keys in Schnorr \citep{Schnorr91}-like signatures (e.g., ECDSA, SchnorrQ) through pre-computing them and storing their hash commitments on the verifier. Despite their merits, these schemes are also vulnerable to quantum computers as they rely on traditional intractability assumptions (e.g., (EC)DLP). \vspace{1mm}

%%%%%%%%%%%%%%%%%%%%%%%%%%%%%%%%%%%%%%%%%%%%%%%%%%%%%%%%%% PQ-secure Signatures %%%%%%%%%%%%%%%%%%%%%%%%%%%%%%%%%%%%%%%%%%%%%%%%%%%%%%%%%%
\noindent{\em \textbf{PQ-secure Signatures}}:  The NIST-PQC standardization \citep{NIST2022selected,darzi2023envisioning} features $\lbbased$ signatures CRYSTALS-Dilithium \citep{Dilithium} and Falcon \citep{fouque2019fast}, alongside $\hbbased$ SPHINCS+ \citep{SPHINCSPLUS} for PQ-secure signatures. Dilithium, based on the Fiat-Shamir with Aborts principle \citep{lyubashevsky2009fiat} and M-LWE and M-SIS problems \citep{brakerski2014leveled, langlois2015worst}, uses a uniform distribution to reduce the public key size by about half, although it results in slightly larger signatures and its variable signing time may impact performance on resource-limited devices. Falcon, a hash-and-sign scheme \citep{ducas2016fast} using NTRU lattices \citep{ducas2014efficient}, requires double-precision floating-point arithmetic and complex operations like Fast Fourier Transform and matrix decomposition, making it unsuitable for devices lacking a Floating-point Unit (FPU). While for 128-bit security, Falcon's public key and signature sizes (897 and 690 Bytes, respectively) are smaller than Dilithium's (1.28 KB and 2.36 KB, respectively), both are less appropriate for resource-limited devices due to their complexity, larger key and signature sizes, and vulnerability to side-channel attacks \citep{Tachyon, courtois2001achieve, NTRU:BLISS:Ducas:Crypto:2013}. Currently, there are no deployable open-source implementations of $\lbbased$ digital signatures for highly resource-limited embedded devices (e.g., with 8-bit microprocessors), aside from BLISS \citep{BLISS}, which was not selected as NIST-PQC standard due to heavy side-channel attack vulnerabilities \citep{tibouchi2020one}. \vspace{1mm}

%%%%%%%%%%%%%%%%%%%%%%%%%%%%%% Lightweight PQ-secure Signatures with Special Assumptions %%%%%%%%%%%%%%%%%%%%%%%%%%%%%%
\noindent{\em \textbf{Lightweight PQ-secure Signatures with Special Assumptions}}: Recent lightweight PQ-secure signature schemes rely on specific architectural features and distributed computation to enhance the signer efficiency. ANT \citep{ANT:ACSAC:2021} transforms a $\lbbased$ OTS into a polynomially-bounded many-time signature through distributed public key computation under the assumption of honest-but-curious servers and passive adversaries. HASES \citep{HASES} and its extension \citep{HASESACMTOMM} convert HORS signature \citep{HORS} into a many-time signature, assuming a secure enclave (e.g., Intel SGX \citep{Intel:SGX:2016}) to store the private key, enabling public key derivation and forward security. Despite their signer efficiency, these special assumptions on the verifier side may limit the applicability of these solutions in some NextG IoT settings. \vspace{1mm}

%%%%%%%%%%%%%%%%%%%%%%%%%%%%%%%%%%%%%%%%%%%%%%%%%%%%%%%%%% Hash-based Signatures as the Building Block %%%%%%%%%%%%%%%%%%%%%%%%%%%%%%%%%%%%%%%%%%%%%%%%%%%%%%%%%%
\noindent {\em \textbf{Hash-based Signatures and their Building Blocks}}: Unlike $\lbbased$ schemes, $\hbbased$ standards such as XMSS \citep{XMSSRFC} and SPHINCS+ \citep{SPHINCSPLUS} rely on minimal intractability and number-theoretical assumptions, offering strong PQ security. These schemes use cryptographic hash functions like SHA-256, which are widely available, facilitating the transition to PQ secure options. While some $\hbbased$ signatures provide efficient signing and verification for a limited number of signatures \citep{HORS, HORSIC, Winternitz:Sec:buchmann2011security}, others, such as \citep{SPHINCS:Bernstein:2015, SPHINCS_ARM, XMSS:Buchmann:2011}, offer high security for extended usage but involve large signature sizes and costly signing processes. 

$\hbbased$ FTSs, built on OTSs \citep{Lamport79, MerkleOTS, Winternitz:Sec:buchmann2011security, hulsing2013optimal}, allow a few signature generations with the same key pair, though security diminishes with each additional message. The first FTS, BiBa \citep{BiBa}, prioritized fast verification and small signature sizes but had trade-offs in signing time and public key size. Subsequent FTSs, like HORS \citep{HORS}, built on Lamport OTS with Bos-Chaum signatures and cover-free families, and its variants \citep{SPHINCS:Bernstein:2015, lee2021horsic+, HORSIC, HORSE, TVHORS, INFHORS}, and others like PORS \citep{aumasson2018improving} and FORS \citep{SPHINCSPLUS} enhanced the robustness against weak message attacks \citep{HORSWeakMessageAttack}. However, they have large private keys (e.g., HORSE \citep{HORSE}) and signing times (e.g., HORSIC+ \citep{lee2021horsic+}), or are time-valid secure (e.g., TV-HORS \citep{TVHORS}).

Stateless $\hbbased$ schemes like SPHINCS \citep{SPHINCS} and SPHINCS+ \citep{SPHINCSPLUS} use a hypertree structure to extend FTSs to full-time signatures. This structure involves Merkle trees with Winternitz OTS \citep{Winternitz:Sec:buchmann2011security} as leaves, where each node signs $2^h$ child nodes, with $h$ as the tree height. For example, with $h=50$, a single key can generate one million signatures per second over 30 years. The hypertree's leaves are FTSs (e.g., HORST instances \citep{SPHINCS}), which allow multiple message signing, enhance path collision resilience, and reduce tree height. While this approach decreases signature size using fewer OTS instances, it increases signing time due to Merkle tree generation. Despite implementations for resource-limited devices (e.g., Armed-SPHINCS \citep{ArmedSPHINCS:Hulsing:PKC:2016}), these schemes are computationally intensive, with SPHINCS+ being 330$\times$ and 21$\times$ slower than ECDSA and Dilithium, and its signatures being 362$\times$ and 7$\times$ larger, respectively. Moreover, stateful $\hbbased$ signatures, such as XMSS$^{MT}$ \citep{XMSSmt} and LMS \citep{LMS}, provide strong security and features like forward security. They use a Merkle tree to group $2^h$ OTS key pairs into one signature key pair, with the root as the public key and each leaf as an OTS. A Merkle signature includes the OTS key pair index, the OTS signature, and the authentication path. However, their high computational cost and memory usage limit their practicality for resource-constrained devices. For example, XMSS$^{MT}$ signatures can be 4.85 KB, which is 105$\times$ and 2$\times$ larger than ECDSA and Dilithium signatures, respectively, and the signing is 430$\times$ and 30$\times$ respectively slower.

Among $\hbbased$ signature schemes, HORS~\citep{HORS} is valued for its efficiency and underpins signatures like XMSS$^{MT}$ \citep{XMSS} and SPHINCS+ \citep{SPHINCSPLUS}. However, extending HORS FTS to full-time signatures using hypertree-based (e.g., SPHINCS+) or tree-based (e.g., XMSS$^{MT}$) methods is inefficient for resource-constrained devices. An alternative is the online/offline paradigm (as in \citep{SEMECS}), where public keys are pre-computed and stored offline, with one key used per signing round. While HORS benefits from small $k$ values for short signatures, it requires large $t$ for adequate security (e.g., 128-bit). For example, signing $2^{20}$ messages with 128-bit security requires $t=1024$ and $k=25$, leading to 32GB of public key storage. In each signing round, 25 keys are used and 999 discarded, resulting in a 97\% key loss and only 2.44\% effective utilization of the public key storage. This inefficiency affects device utility, authentication lifetime, and computation/storage requirements. To reduce key discards in HORS, private keys can be tracked via indices (one per key), with each $t$ key derived from the master key padded with the corresponding index. However, this memory-inefficient approach requires 4KB of storage for $t=1024$ (assuming each index occupies 4 bytes). For microcontrollers like the ATmega328, which have limited flash memory and a threshold of 10,000 write/erase cycles, maintaining the index list in SRAM reduces flash write operations and preserves the data without corruption. Despite improving HORSE's \citep{HORSE} computationally expensive hash-based key generation (using Jakobsson hash calls \citep{jakobsson2002fractal}), this method still risks depleting public key chains (on verifier), potentially causing early termination.

\section{Conclusion}

As next-generation networks increasingly rely on heterogeneous IoTs with resource-limited devices, ensuring secure, scalable authentication is paramount. While traditional digital signatures provide necessary authentication tools, they face significant challenges due to the emerging threat of quantum computing and the inadequacies of current post-quantum cryptography (PQC) solutions for resource-constrained devices. To address these limitations, our Maximum Utilization Multiple HORS (\mumhors) signature scheme offers a lightweight, PQ-secure alternative tailored for IoT applications. With its ability to deliver fast signing, short signatures, and efficient key utilization, \mumhors~provides significant improvements over existing solutions. The experimental results show its superior performance on embedded platforms, making it a promising candidate for secure, resource-efficient digital signatures in IoTs. Hence, \mumhors~paves the way for more secure, scalable, and practical cryptographic solutions tailored to the evolving needs of applications like wearable medical devices or digital twins.

\section{Acknowledgement}
This research is partially supported by the Cisco Research Award (220159).

{\bibliographystyle{plain}
{\bibliography{ref,ref2,AttilaYavuz,crypto-etc,kiarash,SaifNouma}}

\appendix
\newpage 
\section{Bitmap Functionalities Using Linked List} \label{sec:appendix}
This section details the implementation of \bitmap~functionalities using a linked list instead of a circular queue and examines the resulting performance changes. The same design applies to managing public keys on the verifier. This design change affects only the key generation phase of \mumhors~while signing and verification proceed unchanged. Consequently, we present the key generation for \mumhors~and the full details of the new implementation of the \bitmap.

\subsection{MUM-HORS Key Generation and Bitmap Operations}
Based on Algorithm \ref{alg:mumhorsv2}, key generation changes affect Steps 2 and 3. In Step 2, \bitmap~is initialized as a linked list with $rt$ nodes, where $head$ and $tail$ are pointers to the first and last nodes, and each row has a $next$ parameter pointing to the subsequent row. In Step 3, public keys are generated similarly as a $r$-node linked list, with each node containing $t$ public keys derived from the master key $msk$. Note that the verifier does not need to load the entire public key list into memory to avoid overhead. Instead, public keys are stored on disk, and only $rt$ nodes are loaded into memory as needed. Both signer and verifier share the semantics of the \bitmap~parameters for synchronization as before.

\begin{algorithm}
    \footnotesize
    \caption{Key Generation of MUM-HORS (Linked List Version)}\label{alg:mumhorsv2}
    \hspace{5pt}
     
    %%%%%%%% Keygen %%%%%%%%
    \begin{algorithmic}[1]
        \vspace{-5pt}
        \Statex $ \underline{ (\sk, \pk, \bitmap, I_{\mumhors}) \as \mumhorskg(1^{\kappa}) } $:
        \vspace{-1pt}
        \State Set $I_{\mumhors} \as (I_{\hors}, r, rt)$, $msk \Ra \{0,1\}^\kappa$, and $\{pad_{i}\}_{i=1}^3$
        
        \State Create bitmap $\bitmap = \{row_i\}_{i=1}^{rt}$ and set $row_i.num = i$, $row_i.activebits = t$, $row_i.next = row_{i+1}$, all $row_i.bits[.]$ to 1, and $row_{rt}.next = \bot$. Set $\bitmap.head = row_1$, $\bitmap.tail = row_{rt}$, $\bitmap.window = t$, $\bitmap.nextrow = rt+1$, $\bitmap.activerows = rt$, and $\bitmap.activebits = rt\times t$. 

        \State Create public keys $\pk \!\!\as\!\! \{pk_{i}\}_{i=1}^r$ and set $pk_i.num = i$, $pk_i.activepks = t$, $pk_i.next = pk_{i+1}$, $\{pk_i.keys[j] \!\!\!\as\!\!\! f(PRF(msk || i || j))\}_{\substack{1 \leq i \leq r , 1 \leq j \leq t}}$, and $pk_{r}.next = \bot$. Set $\pk.head = pk_1$, $\pk.tail = pk_{rt}$, $\pk.window = t$, $\pk.nextrow = rt+1$, $\pk.activerows = rt$, and $\pk.activepks = rt\times t$.

        \State Send the private key $\sk \as msk$, $\{pad_{i}\}_{i=1}^3$, and $\bitmap$ to the signer, store the public keys $\pk$ and $\{pad_{i}\}_{i=1}^3$ on the verifier, and output the $I_{\mumhors}$
    \end{algorithmic}
\end{algorithm}

The \bitmap~operations using a linked list are detailed in Algorithm \ref{alg:bitmaplist}. The function $\bitmapextendmatrix(.)$ extends the \bitmap~if there are insufficient set bits (less than $window$ bits), calling $\bitmapcleanup(.)$ to remove rows with zero active bits or, if necessary, the row with the fewest active bits. Fresh rows are then added to the end of the list, which is pointed to by the $tail$ parameter. The $\bitmapgetrowcol(.)$ function returns the row and column indices of the $(index+1)^{th}$ set bit for the given $index$ by traversing the list instead of a circular queue. The $\bitmapunsetindices(.)$ function unsets the provided indices where for each index $index$, the $(index+1)^{th}$ set bit of the window is set to zero, with the main difference being the traversal method.

\begin{algorithm}[H]
    \footnotesize
    \caption{Bitmap Manipulation Operations (Linked List Version)}\label{alg:bitmaplist}
    \hspace{5pt}
    %%%%%%%%%% BITMAP EXTEND %%%%%%%%%%
    \begin{algorithmic}[1]
        \vspace{-5pt}
	\Statex $b \as \underline{\bitmapextendmatrix (\bitmap)}$:
        \If{$\bitmap.activebits < \bitmap.window$}
            \If{$\bitmap.nextrow > r$} \Return $\smallfalse$ \EndIf
            \If {$\bitmapcleanup(\bitmap) == 0$}  %\Comment{No row was removed, so choose the best }
                \State Set $row$ to the row with the minimum $activebits$ parameter
                \State Update $\bitmap.activerows \minusEquals 1$ and $\bitmap.activebits \minusEquals row.activebits$
                \State Remove $row$ from $\bitmap$
            \EndIf
            \State $fillcapacity = min(rt - \bitmap.activerows, r - \bitmap.nextrow)$
            \State Add $fillcapacity$ rows to $\bitmap$ by creating new $row$ and setting $row.num = \bitmap.nextrow$, $row.activebits = t$, all $row.bits[.]$ to 1, $row.next = \bot$, $\bitmap.tail.next = row$ and $\bitmap.nextrow \plusEquals 1$ for each.
        \EndIf
            \State \Return $\smalltrue$ 

    \end{algorithmic}
    \algrule

    %%%%%%%%%% BITMAP CLEANUP %%%%%%%%%%
    \begin{algorithmic}[1]
	\Statex $n \as \underline{\bitmapcleanup(\bitmap)}$: Set $cleaned = 0$ and $row = \bitmap.head$
        \While{$row$ is not $\bot$} 
            \If{$row.activebits$ == 0} \State Remove $row$ from $\bitmap$ and set $cleaned \mathrel{+}= 1$ and $row = row.next$ \EndIf
        \EndWhile
        \State \Return $cleaned$
    \end{algorithmic}
    \algrule
    %%%%%%%%%% BITMAP Get Row and Col %%%%%%%%%%
    \begin{algorithmic}[1]
        \Statex $(row, col) \as \underline{\bitmapgetrowcol(\bitmap, index)}$: Set $row = \bitmap.head$
        \While{$row$ is not $\bot$}
            \If{$index < row.activebits$} \textbf{break} \EndIf
            \State Update $index \minusEquals row.activebits$ and $row = row.next$
        \EndWhile
        \State Set $colidx$ as the index of $(index+1)^{th}$ set bit in $row.bits[.]$ and \textbf{return} $(row.num, colidx)$
    \end{algorithmic}
    \algrule
    
    \begin{algorithmic}[1]
        \Statex $\underline{\bitmapunsetindices(\bitmap, \indices)}$: 
        % Set $delta = 0$
        % \State Sort $\indices$ in descending order % and \textbf{if} {$\indices[i] \neq \indices[i-1]$} \textbf{then} $\indices[i] \minusEquals delta$ and $delta \plusEquals 1$ 
        
        \For{$index$ \textbf{in} $\indices$}
            \State $row = \bitmap.head$
            \While{$row$ is not $\bot$}
                \If{$index < row.activebits$} \textbf{break} \EndIf
                \State $index  \minusEquals row.activebits$ {and} $row = row.next$
            \EndWhile

            \State Unset the $(index+1)^{th}$ bit in $row.bits[.]$ to 0
            \State Update $\bitmap.activebits \minusEquals 1$ {and} $row.activebits \minusEquals 1$
        \EndFor  
    \end{algorithmic}
\end{algorithm}

\subsection{Performance Evaluation and Comparison}
{Using} a linked list instead of a circular queue adds an 8-byte $next$ field per node, increasing memory usage by 88 bytes for ($rt=11$) rows on a 64-bit system. Although linked lists enable faster iteration compared to shifting rows in a circular queue—particularly for middle rows—the additional memory overhead is a tradeoff for reduced computational complexity. On a commodity hardware $\bitmapcleanup(.)$ takes 0.027$\mu s$, $\bitmapunsetindices(.)$ 0.18$\mu s$ per index, and $\bitmapgetrowcol(.)$ 0.23$\mu s$. On Raspberry Pi 4, the performance is: $\bitmapcleanup(.)$ takes 0.51$\mu s$, $\bitmapunsetindices(.)$ 0.77$\mu s$ per index, and $\bitmapgetrowcol(.)$ 1.47$\mu s$. The array-based implementation incurs additional computation due to row shifting and slightly higher costs for row iteration due to being a circular queue.

\section{\mumhors~Signer and Verifier Index Synchronization}\label{sec:appendix-sca}
Given the challenge of state synchronization in MTSs built on OTSs \citep{mcgrew2016state}, \mumhors~is vulnerable to desynchronization if messages are corrupted in transit. To address this, we propose the Second Chance Algorithm (SCA), allowing the verifier to recover from out-of-sync states by retaining mismatched public keys. This self-correcting approach enables the verifier to restore synchronization using future valid message-signature pairs without requiring communication with the signer. The new \mumhors~verifier is given in Algorithm \ref{alg:mumhorsverfier2}.

The first two steps and steps 11-12 are similar to the \mumhors's~verifier presented in Algorithm \ref{alg:mumhors}. SCA algorithm allows each rejected public key to remain in the list for a second chance, as the rejection cause (message or signature corruption) may be unclear. Normal public keys are denoted as $pk_{ij}$ and those given a second chance as $pk^*_{ij}$. Using the property of HORS key indices (distance between each key), we apply the following rules: (i) For a verified pair $(s_i, s_j)$ with $i < j$, if there are $j-i$ public keys between them, mark all $pk^*_{ij}$ in between as $pk_{ij}$. (ii) For a verified pair $(s_i, s_{i+1})$, mark all $pk^*_{ij}$ in between as $\bot$ (deleted). If a signature cannot be validated, we not only mark the $pk_{ij}$ as $pk^*_{ij}$ but also delete the actual value of the public key for the received private key and mark it as $\bot$ if the signature is not corrupted (Steps 3-10).

It is important to note that, according to the security proof in Section \ref{sec:security}, the advantage in forging a signature does not arise from corrupting the signature along with the message, as its security relies on the pre-image resistance of the one-way function used. Therefore, we can assume that any signature corruption results from transmission errors, which can be corrected using error correction codes. As a result, the SCA algorithm can be optimized under the assumption of message corruption. However, the strict rules of the SCA algorithm may hinder immediate state recovery during many consecutive corruptions, as the number of corruptions could exceed $j-j'$ between two indices $i_j$ and $i_{j'}$.

The SCA algorithm provides a probabilistic solution to the synchronization issue. However, the most reliable approach is to reset both the signer and verifier to a fresh row number—flushing the bitmap and verifier's memory—once the number of corruptions exceeds a specified threshold. This can be managed through an additional communication mechanism and software checks. {\em It is worthy to note this fact again that the synchronization issue is inherent to stateful $\hbbased$ signatures and is not only limited to our scheme \citep{mcgrew2016state}.}

\begin{algorithm}[t]
    \footnotesize
    \caption{MUM-HORS Verifier with SCA algorithm}\label{alg:mumhorsverfier2}
    \hspace{5pt}

    %%%%%%%% Verify %%%%%%%%
    \begin{algorithmic}[1]
        \vspace{-5pt}
        \Statex $ \underline{ b \as \mumhorsver( \pk, m, \sigma)}$: Set $hash \as Trunc(H(m), k\log t)$ and $b=1$
        \vspace{3pt}

        \State \textbf{if} $hash$ or any $pad_{1,2,3}$ yield distinct $i_j$ as steps 1-2 of $\mumhorssig(.)$, \textbf{then} \textbf{goto} 3 

        \State Split $hash' \as Trunc(H(hash || Ctr), k\log t)$ as step 1 of $\mumhorssig(.)$ into integer indices $i_j$. \textbf{if} {$\exists p,q \in [1, k]$ s.t. $i_p = i_q$ and $p\neq q$} \textbf{then} Execute steps 3-4 of $\mumhorssig(.)$ and \textbf{return} $b=0$.
    
        \For{$j=1$ to $k$}
           \State Find the $(i_j)^{th}$ public key as in $\bitmapgetrowcol(.)$ and count all the $pk^*_{ij}$ until that point as $doubt$
            \If{$s_j$ is verified} mark the corresponding public key as $\bot$
            \Else
                \State Mark the $s_j$'s public key as $pk^*_{ij}$ and try the next $doubt$ non-deleted public keys. 
                \If {any verified}
                    check with last verified signature ($s_{j'}$ given $j>j'$). \textbf{if} there are $j-j'$ public keys \textbf{then} mark all the $pk^*_{ij}$ between $s_{j'}$ and $s_j$ as $pk_{ij}$, \textbf{else} if $j'=j+1$, then, mark all the the $pk^*_{ij}$ as the $\bot$.
                \Else
                    \hspace{1mm} \Return $b=0$
                \EndIf
            \EndIf    
        \EndFor
        \State \Return $b$

        %%%%%%%%%%%%%%%%%%%%%%%%%%%%%%%
        \algsep{\textit{Verifier is Idle}}
        
        \State Invalidate all the $pk_j$ corresponding to the derived $i_j$ as in $\bitmapunsetindices(.)$
    
        \State Extend the view of the public keys similar to $\bitmapextendmatrix(.)$
    
    \end{algorithmic}
\end{algorithm}

\end{document}